\documentclass[twocolumn]{aastex61}
\usepackage{graphicx}
\usepackage{longtable}
\usepackage{threeparttable}
\usepackage{subfigure}
\usepackage[T1]{fontenc}

\newcommand{\Chandra}{\textit{Chandra}}

\newcommand{\NuSTAR}{\textit{NuSTAR}}
\newcommand{\XMM}{\textit{XMM-Newton}}
\newcommand{\swift}{\textit{Swift-XRT}}
\newcommand{\CXOJ}{CXO\,J005215.4--731915}

\begin{document}
\title{Neutron Stars and Black Holes in the Small Magellanic Cloud: The SMC \NuSTAR\ Legacy Survey}
\author{M. Lazzarini}
\affiliation{Department of Astronomy, Box 351580, University of Washington, Seattle, WA 98195, USA}
\author{B. F. Williams}
\affiliation{Department of Astronomy, Box 351580, University of Washington, Seattle, WA 98195, USA}
\author{A. E. Hornschemeier}
\affiliation{Laboratory for X-ray Astrophysics, NASA Goddard Space Flight Center, Code 662, Greenbelt, MD 20771, USA}
\affiliation{Department of Physics and Astronomy, Johns Hopkins University, 3400 N. Charles Street, Baltimore, MD 21218, USA}
\author{V. Antoniou}
\affiliation{Center for Astrophysics | Harvard \& Smithsonian, Cambridge, MA 02138, USA}
\affiliation{Department of Physics \& Astronomy, Box 41051, Science Building, Texas Tech University, Lubbock, TX 79409, USA}
\author{G. Vasilopoulos}
\affiliation{Department of Astronomy, Yale University, PO Box 208101, New Haven, CT 06520-8101, USA}
\author{F. Haberl}
\affiliation{Max-Planck-Institut fŸr extraterrestrische Physik, Giessenbachstrasse 1, 85748 Garching, Germany}
\author{N. Vulic}
\affiliation{Laboratory for X-ray Astrophysics, NASA Goddard Space Flight Center, Code 662, Greenbelt, MD 20771, USA}
\affiliation{Department of Astronomy and Center for Space Science and Technology (CRESST), University of Maryland, College Park, MD 20742, USA}
\author{M. Yukita}
\affiliation{Laboratory for X-ray Astrophysics, NASA Goddard Space Flight Center, Code 662, Greenbelt, MD 20771, USA}
\author{A. Zezas}
\affiliation{Physics Department and Institute of Theoretical and Computational Physics, University of Crete, 71003 Heraklion, Crete, Greece; Foundation for Research and Technology-Hellas, 71110 Heraklion, Crete, Greece}
\affiliation{Center for Astrophysics | Harvard \& Smithsonian, Cambridge, MA 02138, USA}
\author{A. Bodaghee}
\affiliation{Department of Chemistry, Physics and Astronomy, Georgia College and State University, Milledgeville, GA 31061, USA}
\author{B. D. Lehmer}
\affiliation{Department of Physics, University of Arkansas, 226 Physics Building, 825 West Dickson Street, Fayetteville, AR 72701, USA}
\author{T. J. Maccarone}
\affiliation{Department of Physics \& Astronomy, Box 41051, Science Building, Texas Tech University, Lubbock, TX 79409, USA}
\author{A. Ptak}
\affiliation{Laboratory for X-ray Astrophysics, NASA Goddard Space Flight Center, Code 662, Greenbelt, MD 20771, USA}
\affiliation{Department of Physics and Astronomy, Johns Hopkins University, 3400 N. Charles Street, Baltimore, MD 21218, USA}
\author{D. Wik}
\affiliation{Department of Physics and Astronomy, University of Utah, 201 James Fletcher Bldg., Salt Lake City, UT 84112, USA}
\affiliation{Laboratory for X-ray Astrophysics, NASA Goddard Space Flight Center, Code 662, Greenbelt, MD 20771, USA}
\author{F. M. Fornasini}
\affiliation{Center for Astrophysics | Harvard \& Smithsonian, Cambridge, MA 02138, USA}
\author{Jaesub Hong}
\affiliation{Center for Astrophysics | Harvard \& Smithsonian, Cambridge, MA 02138, USA}
\author{J. A. Kennea}
\affiliation{Department of Astronomy and Astrophysics, The Pennsylvania State University, University Park, PA 16802, USA}
\author{J. A. Tomsick}
\affiliation{Space Sciences Laboratory, 7 Gauss Way, University of California, Berkeley, CA 94720-7450, USA}
\author{T. Venters}
\affiliation{Laboratory for Astroparticle Physics, NASA Goddard Space Flight Center, Code 661, Greenbelt, MD 20771, USA}
\author{A. Udalski}
\affiliation{Astronomical Observatory, University of Warsaw, Aleje Ujazdowskie 4, 00-478 Warsaw, Poland}
\author{A. Cassity}
\affiliation{Department of Astronomy, Yale University, PO Box 208101, New Haven, CT 06520-8101, USA}
\affiliation{Smith College, Northampton, MA  01063, USA}


\begin{abstract}
We present a source catalog from the first deep hard X-ray ($E>10$ keV) survey of the Small Magellanic Cloud (SMC), the \NuSTAR\ Legacy Survey of the SMC. We observed three fields, for a total exposure time of 1 Ms, along the bar of this nearby star-forming galaxy. Fields were chosen for their young stellar and accreting binary populations. We detected 10 sources above a 3$\sigma$ significance level (4--25 keV) and obtained upper limits on an additional 40 sources. We reached a 3$\sigma$ limiting luminosity in the 4--25 keV band of $\sim$ $10^{35}$ erg s$^{-1}$, allowing us to probe fainter X-ray binary (XRB) populations than has been possible with other extragalactic \NuSTAR\ surveys. We used hard X-ray colors and luminosities to constrain the compact-object type, exploiting the spectral differences between accreting black holes and neutron stars at $E>10$ keV. Several of our sources demonstrate variability consistent with previously observed behavior. We confirmed pulsations for seven pulsars in our 3$\sigma$ sample. We present the first detection of pulsations from a Be-XRB,  SXP305 (CXO J005215.4$-$73191), with an X-ray pulse period of $305.69\pm0.16$ seconds and a likely orbital period of $\sim$1160-1180 days. Bright sources ($\gtrsim 5\times 10^{36}$ erg s$^{-1}$) in our sample have compact-object classifications consistent with their previously reported types in the literature. Lower luminosity sources ($\lesssim 5\times 10^{36}$ erg s$^{-1}$) have X-ray colors and luminosities consistent with multiple classifications. We raise questions about possible spectral differences at low luminosity between SMC pulsars and the Galactic pulsars used to create the diagnostic diagrams.
\end{abstract}

\section{Introduction}\label{intro}

Population studies of X-ray binaries (XRBs) probe how the local star forming environment affects the production of black holes (BH) and neutron stars (NS), the endpoints of evolution for massive stars. Nearby galaxies provide the opportunity to combine observations of accreting BH and NS, observable as XRBs, with detailed observations of their local star forming environments. The XRB population depends on the physical properties of their host galaxies including metallicity \citep[e.g.,][]{Basuzych2013,BasuZych2016,Brorby2016}, star formation rate \citep[e.g.,][]{Ranalli2003,Gilfanov2004,Mineo2012,Antoniou2010,Antoniou&Zezas2016,Lehmer2019}, and stellar mass \citep[e.g.,][]{Boroson2011,Zhang2012,Lehmer2010,Lehmer2017,Antoniou2019}.
Stars with masses greater than $\sim$8 $M_{\odot}$ -- those that go on to form NS and BH at the ends of their lives -- have binary fractions of at least 60\% \citep{Sana2012,Duchene&Kraus2013}, making the XRB phase an important evolutionary stage for a large fraction of the massive stellar populations in galaxies. 

Obtaining better constraints on the formation and evolution of XRBs is key to understanding binary star evolution, the creation of binary compact-object systems detectable with gravitational waves, and to understanding the heating of the primordial intergalactic medium (IGM) out of which the first galaxies formed \citep[e.g.,][]{Mesinger,Madau2017,Greig2018}. These topics all require information on the demographics of a population of XRBs (fraction with BH and NS primaries) and their dependence on the metallicity and star formation of the surrounding stellar population. 

Completing a full population study of XRBs in the Milky Way is challenging due to the wide range of distances to these systems and reddening because of dust in the Milky Way disk. There has been some successful work \citep[e.g.,][]{Grimm2002,Bodaghee2012,Lutovinov2013,Sidoli2018}, but it is difficult to survey a whole population down to a low enough $L_{X}$ to observe the quiescent population of XRBs. Recent surveys of the Galactic center and Norma arm with \NuSTAR\ have added to our understanding of the Galactic XRB population at hard X-ray energies \citep{Hong2016,Fornasini2017}. In extragalactic XRB populations all sources are at the same distance, allowing for accurate measurement of source luminosities. 

Previous studies of Local Group galaxies with X-ray missions such as \Chandra\ and \XMM\ have connected the XRB populations with the ages of the stellar populations hosting them \citep[e.g.,][]{Antoniou2009,Antoniou2010,Antoniou2019,Antoniou&Zezas2016,Garofali2018,Lazzarini2018,Williams2018}. However, the soft ($E<10$ keV) X-ray band alone does not allow us to distinguish among the compact-object types for an entire population of XRBs.

With the launch of the \textit{Nuclear Spectroscopic Telescope Array} (\NuSTAR) in 2012 \citep{Harrison2013}, we are now able to use the 4--25 keV energy range to study extragalactic populations \citep[e.g.,][]{Wik2014,Yukita2016,Vulic2018}. An entire population of XRBs can be separated into groups according to compact-object type using \NuSTAR\ because of spectral differences in the hard band ($E<10$ keV). We can distinguish XRBs with BH and NS primaries by comparing their X-ray luminosities and colors with those of Galactic XRBs of known compact-object type (Zezas et al., in preparation).

BH and NS XRBs may be further subdivided into accretion states (BHs) and by magnetic field strength (NS). As the accretion rate of a BH XRB varies, it undergoes spectral state transitions, commonly referred to as accretion states. Its X-ray luminosity and hard colors vary due to shifts in the dominant emission mechanism \citep[for more detailed overview see e.g.,][]{Remillard&McClintock,Done2007,Tetarenko2016}. NS XRBs can also be classified as accreting pulsars (high magnetic field) or low magnetic field neutron stars (Z-type and atoll-type), with these two groups separated in the X-ray intensity/hardness space. Note that accreting pulsars have harder X-ray spectra than hard state BHs in the energy range we study in this paper \citep[e.g.,][]{Reig2011}.

An intriguing sub-class of XRBs are the ultra luminous X-ray sources (ULXs), bright systems with isotropic luminosities that exceed the  Eddington limit for a stellar mass ($\sim$10$-$20 M$_{\odot}$) BH \citep{2017ARA&A..55..303K}.
It was initially suggested that these systems  hosted intermediate-mass BHs accreting at sub-Eddington rates, but has now been established that at least a few of them  host pulsating NSs \citep{2014Natur.514..202B,2018MNRAS.476L..45C,2016ApJ...831L..14F,2017MNRAS.466L..48I}. In addition, it has been shown that the spectral properties of pulsating and non pulsating ULXs share similarities and are consistent with theoretical predictions of super-Eddington accretion onto a NS \citep{2017A&A...608A..47K,2018ApJ...856..128W}. Moreover, the recent discovery of Be-XRB pulsars that have gone through major outbursts \citep[e.g. Swift J0243.6+6124][]{2018ApJ...863....9W} reaching luminosities near or above 10$^{39}$ erg/s have enabled us to investigate the spectral changes of those systems and compare them with ULXs \citep[e.g.][]{2018A&A...614A..23K}. These studies have demonstrated that XRB pulsars that are traditionally thought as some of the harder accreting systems, can become significant softer at high accretion rates, while exhibiting a thermal like cut-off in their X-ray spectra \citep{2017A&A...608A..47K}. Thus ULXs can be used to implement any diagnostic tool developed to classify systems based on their spectral properties.   

A. Zezas et al. (in preparation) have developed a diagnostic for determining compact-object type in extragalactic XRB populations using a sample of Rossi X-ray Timing Explorer (\textit{RXTE}) Proportional Counter Array (PCA) spectra from Galactic XRBs of known compact object type. The hard X-ray coverage of \textit{RXTE} makes these observations comparable to \NuSTAR\ observations in the 4--25 keV energy band, when adjustments are made for instrument response.

The diagnostic diagrams created with X-ray colors and luminosities have been used to classify compact-objects in nearby galaxies including M83 \citep{Yukita2016}, NGC 253 \citep{Lehmer2013,Wik2014}, M33 \citep[][Yang et al. 2019, in preparation]{West2018}, and M31 \citep[][D. Wik et al., 2019, in preparation]{Yukita,Stiele2018,Lazzarini2018}. \citet{Vulic2018} applied this method to a larger sample of 12 galaxies within 5 Mpc. Here we provide \NuSTAR-based classifications of the XRB population in the Small Magellanic Cloud (SMC). Due to the proximity of the SMC \citep[$D=61.9\pm0.6$ kpc;][]{deGrijs2015}, we probe to lower point source luminosities in the 4--25 keV band than other \NuSTAR\ observed galaxies.

The SMC is the second closest star-forming galaxy to the Milky Way \citep{Hilditch2005} with a well-mapped star formation history \citep{Zaritsky&Harris2004,Rubele2018}. Beyond its proximity and depth of study, the SMC is an interesting environment for studying XRBs because it has a comparable number of confirmed and candidate high mass XRBs (HMXBs) to the Milky Way --- $\sim$120 compared to $\sim$110 --- \citep{Haberl2016,Liu2005,Liu2006,Krivonos2012}. 

The low metallicity of the SMC, $Z \sim 1/5\ Z_{\odot}$ \citep[e.g.,][]{Luck1998,Antoniou&Zezas2016}, makes it an interesting comparison point with the XRB populations observed with \NuSTAR\ in other galaxies. Metallicity has been seen to cause variations in the XRB luminosity function \citep{BasuZych2016,Lehmer2019}, with low-metallicity galaxies hosting more luminous HMXBs. \citet{Douna2015} found that low metallicity galaxies hosted roughly 10 times the number of $L>10^{39}$ erg s$^{-1}$ HMXBs seen in solar metallicity galaxies.

Of the HMXBs in the SMC, all but possibly two of the confirmed HMXBs \citep{Maravelias2014} are known to be Be/X-ray binaries \citep[e.g.,][]{Haberl2016}, where the stellar companion is an Oe or Be star. Given the high number of known HMXB systems in the SMC, there is a noticeable absence of systems with confirmed BH accretors \citep{Liu2005}. Actually, there is only one Be/X-ray binary system with a confirmed BH accretor \citep{Casares2014}, and that system is in the Milky Way. \citet{Zhang2004} have proposed that the dearth in observed BH-HMXBs may be because Be/BH binaries are transient systems with a long quiescent state. Another possibility for the scarcity of Be/BH systems is that their formation is disfavored by binary evolution \citep{Belczynski2009}.

Distinguishing between BH and NS XRBs is a challenging problem to which there are currently limited solutions. A NS can be confirmed if a low-mass XRB (LMXB) has a Type IX-ray burst \citep{Type1} or if pulsations are observed. BHs can be classified if their companion star has a well measured orbital period, radial velocity amplitude and constrained inclination angle, al of which allow a constraint to be placed on the compact-object mass \citep{Orosz}. There is of course also the new prospect of precision mass measurements via gravitational waves that can give estimations of masses indicative of NS versus BH \citep[e.g.,][]{Abbot2016,LIGO_neutronstars,LIGO_catalog}; however, gravitational waves can only be detected by the LIGO detectors after the XRB phase has ended.

In this paper we present deep \NuSTAR\ observations (1 Ms in total) of three fields along the SMC bar chosen to maximize the number of observed HMXBs. We present source classifications for selected sources with well constrained X-ray luminosities and hardness ratios. In Section \ref{analysis}, we describe the \NuSTAR\ observations used in this work and we describe the data reduction methods used. In Section \ref{classification}, we discuss how compact-objects were classified using their X-ray luminosities and hardness ratios. In Section \ref{results}, we present our results and discuss individual sources of interest. In Section \ref{conclusions}, we present a brief summary of our results.

\begin{figure*}
\centering
\includegraphics[width=0.99\textwidth]{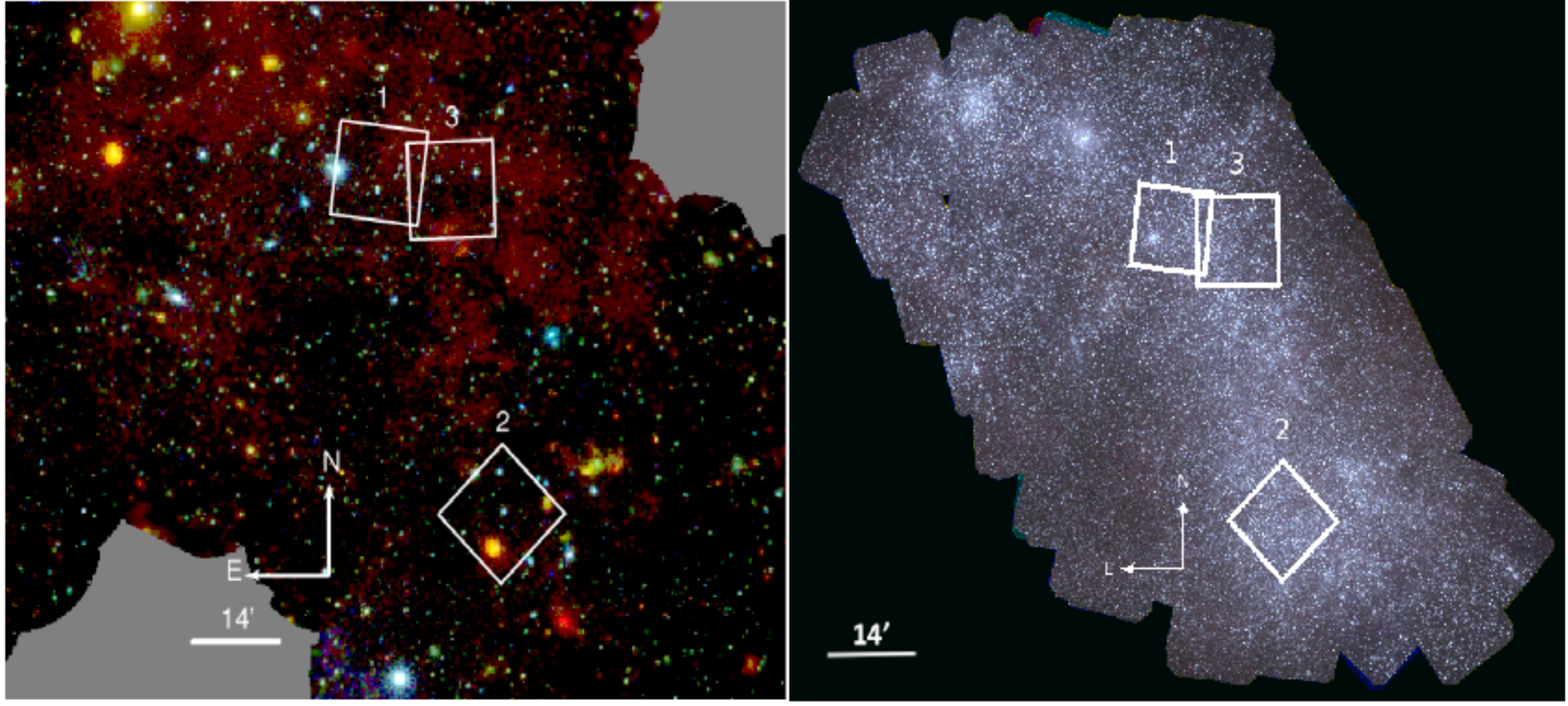}
\caption{\XMM\ X-ray mosaic image (left) and UV-optical mosaic image (right) of the SMC with the three fields observed by \NuSTAR\ presented in this work marked. Left: The image was created by combining \XMM\ observations in the direction of the nearby star-forming galaxy, available until April 2017 \citep{Maitra2019} and following the procedure described by \citet{Haberl2012}. The RGB color is composed of three energy bands 0.2-1.0 keV (red), 1.0-2.0 keV (green) and 2.0-4.5 keV (blue). Colors of point sources are characteristic of their nature, with orange being mostly SNRs, green galaxy clusters or background AGNs and blue HMXBs. Right: \textit{Swift} ultraviolet optical telescope (UVOT) mosaic image of the Small Magellanic Cloud with the three fields observed by \NuSTAR\ labeled. This RGB image was created using the following filters: blue = \textit{uvw2}, green = \textit{uvm2}, red = \textit{uvw1}. The three \textit{Swift} filters have the following central wavelengths, respectively: 1928, 2246, 2600 \AA\ \citep{Hagen2017}.}
\label{smc}
\end{figure*}

Throughout this work we assume a Galactic neutral hydrogen column density of $6.65 \times 10^{20}$ cm$^{-2}$ for Field 1, $4.53 \times 10^{20}$ cm$^{-2}$ for Field 2, and $6.90 \times 10^{20}$ cm$^{-2}$ for Field 3 \citep[][see Table \ref{obs_list} and Figure \ref{smc} for field locations]{Dickey&Lockman1990} for converting \NuSTAR\ count rates to fluxes. We assume a distance of 61.9$\pm$0.6 kpc to the SMC \citep{deGrijs2015} to convert fluxes to luminosities.

\begin{deluxetable*}{ccccccl}
\tablecaption{Log of \NuSTAR\ Observations\label{obs_list}}
\tablehead{
\colhead{Obs. ID} &  
\colhead{R.A.} & 
\colhead{Dec.} & 
\colhead{Field} & 
\colhead{Exposure} & 
\colhead{Date (start)} & 
\colhead{Notes} \\ 
\colhead{} &  
\colhead{(J2000)} & 
\colhead{(J2000)} & 
\colhead{ID} & 
\colhead{Time [ks]} & 
\colhead{yyyy mon. dd}& 
\colhead{}
}
\startdata
50311001002 & 13.92740 & --72.43900 & 1 & 137 & 2017 Apr 24 & Stray light in FPMB \\
50311001004 & 13.82720 & --72.43990 & 1 & 137 & 2017 Aug 12 &  \\
\textbf{Field 1 Total} & & & & 274 &  \\
\hline
50311002002 & 12.71280 & --73.25750 & 2 & 284 & 2017 Mar 12 & \\
50311002004 & 12.67640 & --73.28180 & 2 & 129 & 2017 Jul 19 &  \\
50311002006 & 12.62280 & --73.27560 & 2 & 46 & 2017 Aug 09 & \\
\textbf{Field 2 Total} & & & & 459 &  \\
\hline 
50311003002 & 13.26350 & --72.48780 & 3 & 146 & 2017 May 03 & Stray light in FPMB \\
50311003004 & 13.17230 & --72.48070 & 3 & 147& 2017 Aug 07 &  \\
\textbf{Field 3 Total} & & & & 293 & \omit \\
\enddata
\tablecomments{Log of \NuSTAR\ observations used in this analysis. More information on stray light contamination can be found in Section \ref{stray_light}. Listed exposure times are combined for FPMA and FPMB telescopes and contain data from good time intervals (see Section \ref{initial_processing} for more details). The total exposure time for the two observations that had stray light contamination in the FPMB images only include the exposure time for FPMA, as the contaminated FPMB images were not included in data analysis.}
\end{deluxetable*}

\section{NuSTAR Data \& Analysis}\label{analysis}
The \NuSTAR\ data were collected over three separate $15^\prime \times 15^\prime$ fields (see Figure \ref{smc}) from 2017 March 12 to 2017 August 12. Fields 1 and 3 were observed in two epochs and field 2 was observed in three epochs. Observations were planned so that each field had a total exposure time of roughly 200 ks, for both focal plane modules A and B (FPMA, FPMB). Table \ref{obs_list} provides an overview of all individual observations and exposure times for each field.

The three fields that comprise this survey were chosen because of their large HMXB populations and potential for hosting elusive BH XRBs. All three fields host young stellar populations that are rich in accreting pulsars. The young stellar populations are likely to host BH XRBs, and there are two HMXBs without detected pulsations, potential BH candidates.
\subsection{Initial Processing}\label{initial_processing}
We reduced the \NuSTAR\ observations using \texttt{HEASOFT v6.24} along with \texttt{CALDB v4.7.9}. We reprocessed Level 1 event files using the \texttt{nupipeline} tool, stopping at Level 2 and using the parameters \texttt{SAAMODE=strict} and \texttt{TENTACLE=yes} to filter out time intervals with high background due to passage through the South Atlantic Anomaly. We used the \texttt{nuproducts} tool to generate light curves for the FPMA and FPMB telescope for each observation. We inspected the light curves to confirm that the observations did not include any flares. 
We generated images with data from good time intervals (GTIs) in the 4-6, 6-12, 12-25 and full 4--25 keV bands using the \texttt{heasoft} tool \texttt{xselect}.
\begin{figure}
    \centering
    \includegraphics[width=0.35\textwidth]{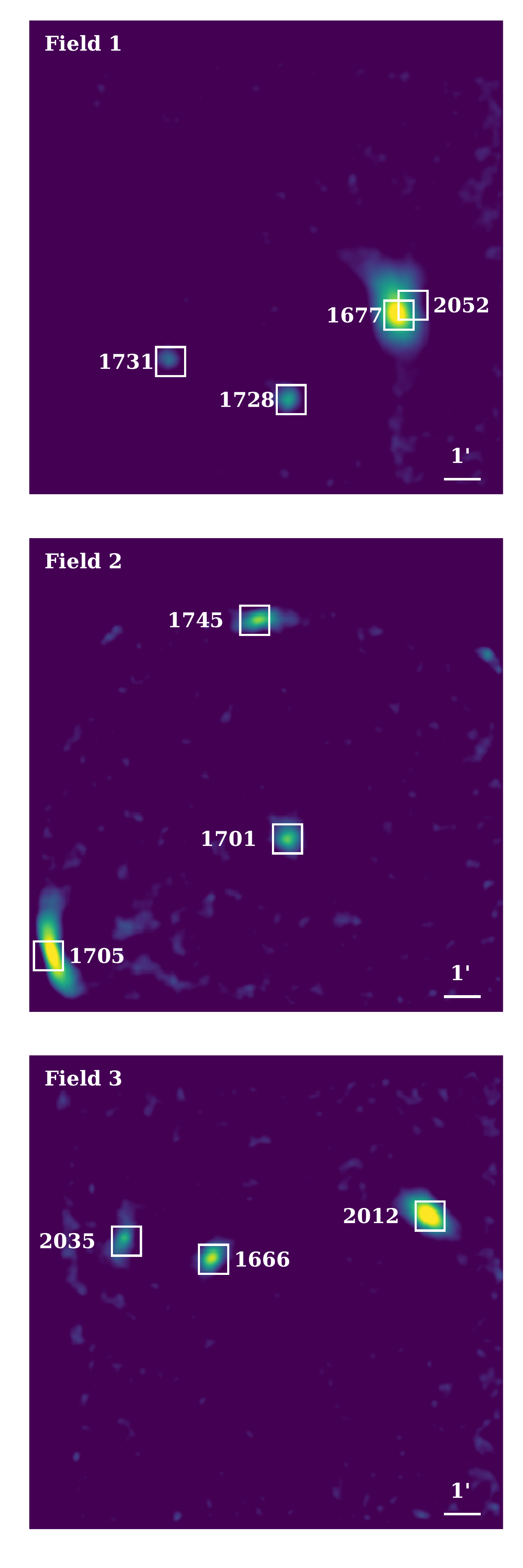}
    \caption{Images of the three \NuSTAR\ fields 1 (top), 2 (center), and 3 (bottom) with sources that are detected above 3$\sigma$ in the 4--25 keV band, marked with white boxes. These images are for display purpose only, not analysis. Images were generated by stacking the 4--25 keV images from each observation for each field. Images were then deconvolved with the \NuSTAR\ PSF from CALDB v4.7.9, using publicly available code by Brian Grefentsette; \small \url{https://github.com/bwgref/m51_deconvolution}.}
    \label{smc_rgb}
\end{figure}

\subsection{Stray Light}\label{stray_light}
We inspected images for stray light contamination. The FPMB telescope images from Field 1 (obsid 50311001002) and Field 3 (obsid 50311003002) both had visible stray light contamination due to the nearby X-ray bright binary SMC X-1. We confirmed the stray light contamination using the publicly available \verb!nustar_stray_light!\footnote{\url{https://github.com/bwgref/nustar_stray_light}} tool. Stray light contamination resulted in a loss of $\sim$45\% of the detector area in obsid 50311001002 and $\sim$40\% of the detector area in obsid 50311003002. Due to the large area lost to stray light contamination and the potential for stray light contamination beyond the regions where stray light is immediately visible by eye, the FPMB images for observations 50311001002 and 50311003002 were omitted from further analysis (background fitting, simultaneous PSF fitting) and are not included in the total exposure times listed in Table \ref{obs_list}.

\subsection{Background Fitting}\label{bgd_fitting}
Background fitting was done using the \verb!nuskybgd! tool \citep{Wik2014}, which is publicly available\footnote{\url{https://github.com/NuSTAR/nuskybgd}}. The background emission in \NuSTAR\ images comes from a combination of stray light from sources outside the field of view (FOV), as well as the cosmic X-ray background, instrumental background, and reflected solar X-rays. The \verb!nuskybgd! tool fits combinations of models of the aforementioned types of emission to extracted background spectra from source-free regions in \NuSTAR\ images, with the aim to model the position and energy dependent background emission to improve our source characterization. Because stray light regions were masked out of our images prior to background fitting, we only fit possible leftover stray light with the background fitting tool.

We fit the background emission in the full 4--25 keV band for each observation and separately for each FPMA and FPMB telescope image, omitting FPMB images for observations 50311001002 and 50311003002. For each module, the \NuSTAR\ FOV is divided between four CCDs (0-3). To account for spatial variation in the background emission across the FOV, we fit each detector separately. We fit individual background models for each observation and the FPMA/FPMB images, which are then applied when fitting for count rates, as described in Section \ref{psffitting}. We follow the methodology used in \citet{Vulic2018}, while for a more detailed overview of the \verb!nuskybgd! tool, see \citet{Wik2014}.

\subsection{Point Source Detection with PSF Fitting}\label{psffitting}
Characterizing point source emission in crowded regions is challenging with \NuSTAR, especially given its relatively broad point spread function (PSF) in comparison to $E<10$ keV imaging telescopes such as \XMM\ and \Chandra. \NuSTAR's PSF core has a full width at half maximum of 18\arcsec\ and a half-power diameter of 58\arcsec\ \citep{Harrison2013}. In crowded regions, emission from point sources can be contaminated by the PSF wings of other nearby sources. To account for this, we fit point source count rates and hardness ratios using simultaneous PSF fitting for an input source catalog, using the method presented in \citet{Wik2014} and following the methodology outlined in \citet{Vulic2018}. The steps of this PSF fitting analysis are described below.

\subsubsection{PSF and Response File Correction}
The \NuSTAR\ telescope distorts the PSF into a banana shape for sources that are off axis ($\theta > 3^{\prime}$) \citep{Harrison2013,Wik2014}. We use the library of \NuSTAR\ PSFs as a function of off-axis angle from the \verb!CALDB! to correct for the shape distortion of source PSFs toward the edges of the FOV. 

To account for energy-dependent vignetting, we generated an energy weighted vignetting function by weighting the \NuSTAR\ vignetting function by a typical XRB power law spectrum. The \NuSTAR\ vignetting function is highly energy-dependent \citep{Harrison2013}, with higher levels of vignetting at high energies. We used this weighted function to generate ancillary response files (ARFs) and created RMFs using the appropriate response file from the \NuSTAR\ \verb!CALDB!.

\subsubsection{Astrometric Alignment}

Astrometric alignment was done via PSF fitting with the input \Chandra\ source catalog of Antoniou et al., in preparation, including detections from the SMC \Chandra\ X-ray Visionary Program survey \citep{Antoniou2019} that observed 11 fields, identified for their young stellar populations, and 3 additional observations from the archive all to a limiting luminosity of $\sim 1.3 \times 10^{32}$ erg s$^{-1}$ in the full (0.5--8.0 keV) band. We chose the brightest 3--4 sources in each field in the 4--25 keV band to calculate the $x$ and $y$ shifts between the \NuSTAR\ and \Chandra\ images. We performed this astrometric alignment independently for each observation and field in our sample with a mean $x$ shift of $\sim1.1$ pixels ($\sim$2.7\arcsec) and mean $y$ shift of $\sim-0.5$ pixels ($\sim$1.2\arcsec). The shifts were then applied when performing PSF fitting in order to estimate the source count rates.

\subsubsection{Count Rate Extraction with Simultaneous PSF Fitting}

PSF fitting was performed within user-defined rectangular regions. We ensured that the edge of each fitting rectangle extended at least $1^{\prime}$ on either side of the input source position to ensure that we exceeded twice the half power diameter of the \NuSTAR\ PSF. When possible, we used one rectangular region to encompass the observation's FOV, but due to the roll angle of certain observations, we used multiple rectangles so that we would eliminate regions outside the FOV from the extraction regions. 

For each rectangular region, we generated the axis-corrected PSFs and vignetting-corrected response files. We generated a background image using the background model produced with \verb!nuskybgd!. Then, a model image was produced by combining the PSFs with the background image. This model image was then fit to the data to extract count rates for each source. For a more detailed discussion of the model fitting procedure, see Section 4.3.2 from \citet{Vulic2018}.

The count rates were fit in soft ($S$; $4-6$ keV), medium ($M$; $6-12$ keV), hard ($H$; $12-25$ keV), and full ($F$; $4-25$ keV) energy bands. These bands were chosen as they provide the most robust separation between types of sources on the diagnostic diagrams \citep[Zezas et al., in preparation;][]{Vulic2018}. We calculated the significance of each source detection using the source count rate, background count rate, and exposure time. The background rate used for each source was determined with the simultaneous PSF fitting code, taking the background model (\S \ref{bgd_fitting}) into account. 

We only use sources with a significance greater than $3\sigma$ in the 4--25 keV band for our source classification analysis, although we report all sources whose positions were input into our PSF fitting routine that returned lower significance measurements as upper limits. We chose the 3$\sigma$ detection threshold because all sources have multi-wavelength counterparts. Specifically, we use the \Chandra\ source positions for all sources from the \citet{Antoniou2019} catalog as priors on our PSF count rate fitting, so we know that all sources have previously been detected at X-ray wavelengths.

The PSF fitting code by \citet{Wik2014} assumes a default photon index of 2. To test the impact of the chosen photon index on our analysis, we ran our analysis with a photon index of 0.9 for sources associated with known pulsars and 1.7 for other sources for a subset of our 3$\sigma$ sample. We found a $\ll 1\%$ difference in the output count rates and hardness ratios when compared to the output with the default photon index, which is expected as the hardness ratios are calculated using count rates rather than fluxes. Given this negligible difference in output, the count rates and hardness ratios reported in this work were all obtained with the default photon index, which allowed for more efficient data analysis. The relatively weak dependence of PSF on energy for \NuSTAR\ may play an important role in this result \citep{Madsen2015}.

In addition to fitting count rates, we used the simultaneous PSF fitting routine to fit the hardness ratios for our sources. We use the technique developed by Wik et al., (in preparation) and described in detail in \citet{Vulic2018}. The hardness ratios we fit were HR1~$ = (M-S)/(M+S)$ and HR2~$ = (H-M)/(H+M)$. We performed simultaneous PSF fitting with the hardness ratios as free parameters. This reduces the errors associated with the hardness ratios because instead of propagating the error on the count rate measurements the HR errors are calculated independently.

We input 50 \Chandra\ source positions for simultaneous PSF fitting with 0.5--8.0 keV luminosities down to $\sim 5 \times 10^{33}$ erg s$^{-1}$ \citep{Antoniou2019}. The luminosity limit for the input \Chandra\ sources was determined by iterating the PSF fitting routine and adding in $\sim$3 \Chandra\ sources in descending luminosity order each time the code was run. We first input only the brightest few \Chandra\ sources that were easily visible in the 4--25 keV \NuSTAR\ images (see Figure \ref{smc_rgb}, and then added about 3--5 sources at a time until additional sources were not detected. We list the positions, count rate in each band, hardness ratios, exposure time, and background count rate for all 50 sources for which we attempted to fit count rates in Table \ref{nustar_table}. These measurements merge all the observations for each field.

We also ran the PSF fitting routine for each individual observation in order to determine source variability and compare with the quasi-simultaneous \swift\ observations. We list the count rates and hardness ratios for each source with greater than 3$\sigma$ significance from each observation in Table \ref{nustar_time}.

\subsection{Using Simultaneous \swift\ Observations to Test PSF Fitting in Crowded Regions}\label{swift_observations}

As part of the SMC \NuSTAR\ Legacy observation program, deep observations of the \NuSTAR\ fields presented in this paper were taken with the Neil Gehrels \swift\ Observatory \citep[\swift\ ,][]{2004ApJ...611.1005G} X-ray Telescope \citep[XRT, ][]{2005SSRv..120..165B} (PI: V. Antoniou). The \swift\ observations were quasi-simultaneous with our \NuSTAR\ observations, taken between 0 and 7 days apart (see Figure \ref{swift_Comparison}). Data were retrieved from the \swift\ data center\footnote{\url{http://www.swift.ac.uk/}}, and were analyzed using standard procedures as outlined in \citet{2007A&A...469..379E,2009MNRAS.397.1177E}, and briefly summarized below.

\swift\ data were reduced using the {\tt xrtpipeline} (v0.13.4), that can be found in  HEASoft 6.23 software\footnote{\citep{HEASOFT} \url{https://heasarc.nasa.gov/lheasoft/}}. Clean events were extracted with the HEASoft FTOOLS \citep{1995ASPC...77..367B}, by using the command line interface {\tt xselect}. Source detection was performed using the command line interface {\tt ximage}. Only sources with significance above 3 $\sigma$ were selected. The complete observing log can be found in Table \ref{obs_list2}. We present the count rates for each source by observation in Table \ref{swift_data}.

\begin{figure*}
    \centering
    \includegraphics[width=0.95\textwidth]{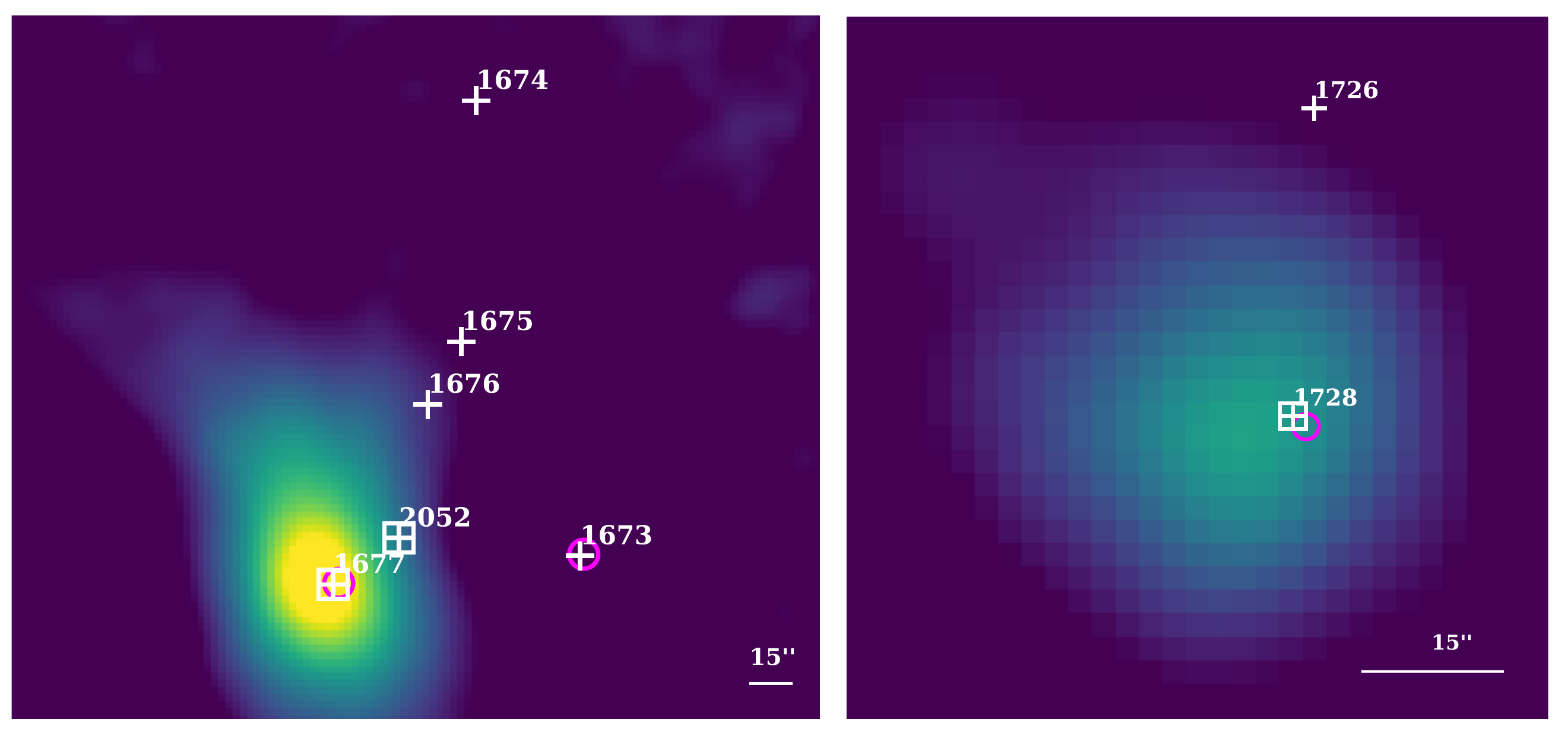}
    \caption{Zoom in of 4--25 keV band deconvolved image of Field 1 shown in Figure \ref{smc_rgb}. Sources for which we obtain upper limits with our \NuSTAR\ observations are marked with white crosses. Sources that were detected by \NuSTAR\ above 3$\sigma$ significance in the 4--25 keV band are marked with a white box and white cross. Magenta circles indicate source detections by \swift. Each circle indicates the average position of the \swift-detected source, weighted by exposure time for each observation. To test how well our PSF fitting code retrieved source count rates and hardness ratios in crowded regions, we first fit for all sources in our input \Chandra\ source catalog with 0.5-8.0 keV luminosities above 5$\times 10^{33}$ erg s$^{-1}$. Next, we only fit for sources that were also detected by \swift\ (marked with magenta circles) and compared the output count rates and hardness ratios. We found that the difference in the measured count rates and hardness ratios for the brightest sources in each region (1677 in the left panel and 1728 in the right panel) shifted by 7\% or less.}
    \label{swift_sim}
\end{figure*}

We used the quasi-simultaneous \swift\ observations to determine how effective our simultaneous PSF fitting code was in extracting count rates for sources in crowded regions. We selected two crowded regions within Field 1 (the regions surrounding sources 1677 and 1728; see Figure \ref{swift_sim}), and performed simultaneous PSF fitting for all \Chandra\ sources with 0.5-8.0 keV flux above 5$\times 10^{33}$ erg s$^{-1}$ within $\sim$0.5$^{\prime}$ of the brightest central source. Then we performed PSF fitting again, only including \Chandra\ sources that had also been detected in the \swift\ observations. 

We found that the measured count rate and hardness ratios for source 1677 did not shift drastically when surrounding sources within 30$^{\prime \prime}$ that were not detected by \swift\ were removed from the input \Chandra\ source list used in PSF fitting. Source 1677 was detected with $\sim$ 342$\sigma$ significance in the full 4--25 keV \NuSTAR\ band. Source 2052 was detected by \NuSTAR\ with $\sim$ 11 $\sigma$ confidence in the full 4--25 keV band, but not detected by \swift. When we removed source 2052 from our input source list and re-performed PSF fitting to extract the count rate for source 1677, we found that the 4--25 keV count rate measured for source 1677 increased by only $\sim 2 \%$. Similarly, when we re-fit for the hardness ratios of source 1677 when source 2052 was removed, we found that both HR1 and HR2 decreased by $\sim 1 \%$ and $\sim 2 \%$, respectively.

We performed the same experiment with source 1728, which was detected by \NuSTAR\ with $\sim$11$\sigma$ significance in the 4--25 keV band, and in the \swift\ observations. Source 1726 is located roughly 30$^{\prime \prime}$ away from source 1728, and was not detected in the \swift\ observations. Source 1726 was not detected with a $\sim$1$\sigma$ upper limit in the 4--25 keV band by \NuSTAR. When we removed source 1726 from our input source list and re-fit for the count rate and hardness ratios for source 1728, we found that the measured 4--25 keV count rate for source 1728 increased by $\sim 7 \%$, while HR1 increased by $\sim 3 \%$ and HR2 decreased by $\sim 4 \%$. 

We found that 4--25 keV count rates and hardness ratios for the brightest sources in the two crowded regions we tested changed by 7\% or less when we omitted input sources that were not detected in the \swift\ observations. We conclude that the PSF-fitting routine was not significantly overfitting the bright sources, and therefore we included all \Chandra\ sources with 0.5-8.0 keV luminosity above $\sim 5 \times 10^{33}$ erg s$^{-1}$ in our input source list to allow the PSF-fitting routine to deconvolve confused sources to the maximum extent possible.

\subsection{\NuSTAR\ Timing Analysis}\label{timing}

We looked for pulsations in the observations of our 10 sources that were detected above 3$\sigma$ significance. We performed analysis for each source using the cleaned combined FPMA and FPMB event list for all observations of each source's field (see Table \ref{obs_list}). For each source we produced a trimmed event file, including all counts within a 20 pixel ($\sim 50^{\prime \prime}$) radius of the source position and selected event corresponding to 4--25 keV photon energies. For the period search we used barycenter corrected event times (barycenter correction was done with the \texttt{barycorr} tool from \texttt{FTOOLS}). 
We performed an epoch folding \citep{1983ApJ...272..256L} test to search for pulsations of each detected system. The test was implemented through \verb!python! by using \verb!stingray! and \verb!HENDRICS! \citep{Huppenkothen2019}. We initially searched all event files for a periodic signal
over a broad range of frequencies from 0.001 - 1 Hz. This was done using the HENzsearch tool. Once a candidate periodic signal was determined, we performed another search within a smaller range around the candidate frequency to get a more precise value, while also fitting a Gaussian curve to the best-fit frequency in order to estimate uncertainties. 
Upon determining a periodic signal we folded the events using the best fit period in order to obtain pulse profiles with 16 bins over a complete pulse phase. 
We then determined the maximum ($R_{\rm M}$) and minimum ($R_{\rm m}$)  values of the pulse profiles and calculated the pulsed fraction as $PF=(R_{\rm M}-R_{\rm m})/(R_{\rm M}+R_{\rm m})$. For systems where no significant period was detected we determined a upper limit for the $PF$ that would have resulted in a 3$\sigma$ detection. 

We were able to confirm pulse periods for all six pulsars in our sample at an above 3$\sigma$ significance level.
Moreover, we detected a pulse period from a candidate HMXB, thus confirming the nature of the compact object.
All period detections had a significance above 3$\sigma$.
Given the long baseline of the \NuSTAR\ observations we have also performed an accelerated epoch folding test to search for period derivative \citep[e.g. see][]{2018A&A...620L..12V}; all period derivatives were consistent with zero ($|\dot{\nu}|<10^{-11}$). 
We note that for source 2052, we were not able to measure a pulse period due to high background emission from the nearby and  much brighter source 1677.

We list the results of our timing analysis for each source in their individual subsections in Section \ref{results}. We present a summary of the pulse periods we measured for each pulsar during each observation along with their published pulse periods in Table \ref{pulse_periods}.

\section{Source classification}\label{classification}
\begin{figure*}
    \centering
    \includegraphics[width=0.95\textwidth,keepaspectratio]{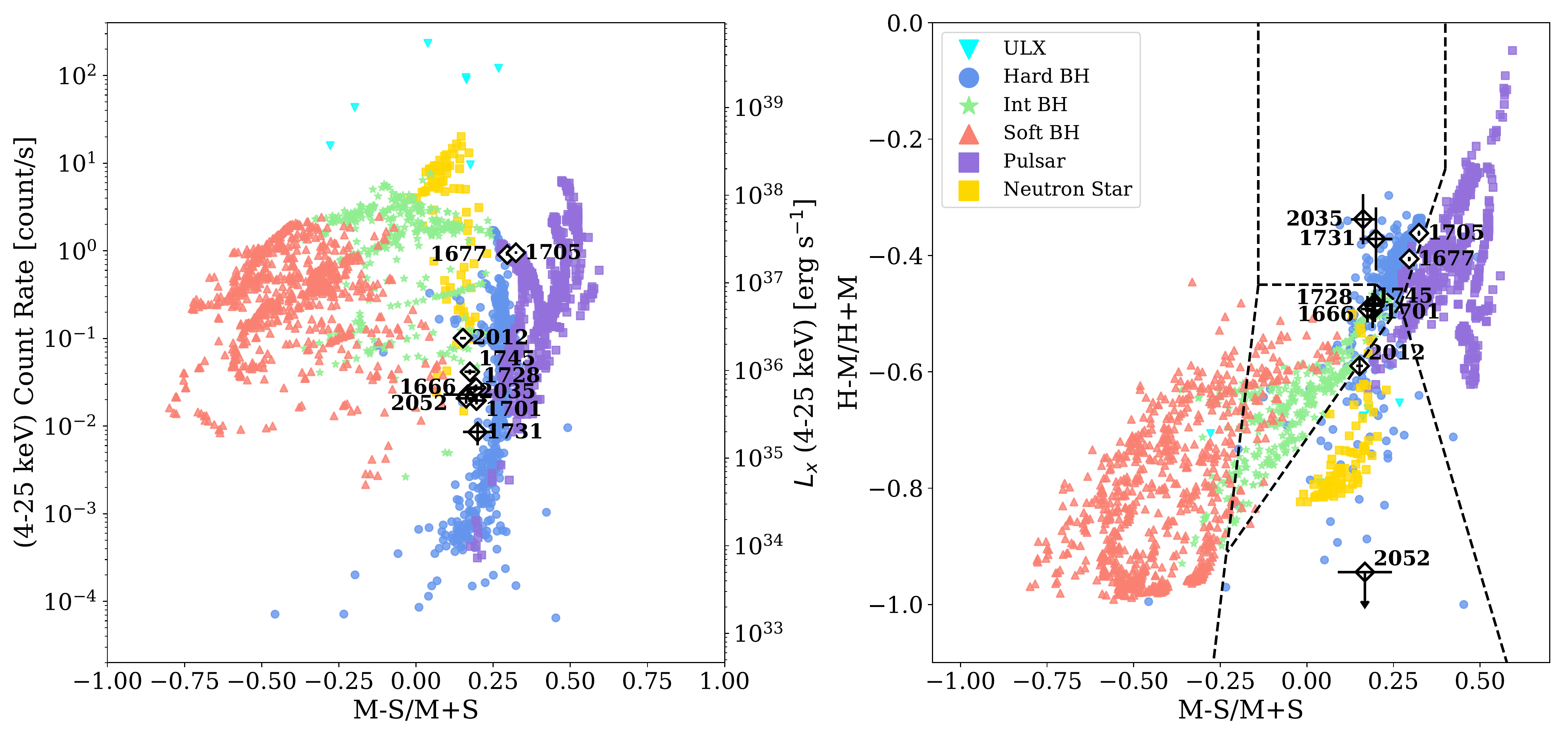}
    \caption{Hardness-intensity diagram and hardness ratio plots for \NuSTAR\ sources combining all epochs of observation for each field. This figure is used to give an overview of the sources in our sample. To classify our sources we used diagnostic diagrams with each epoch of observation plotted separately for each source in order to account for variability between observations (see Figures \ref{var_1666}-\ref{var_2035}). Colored points are Galactic \textit{RXTE-PCA} observations of accreting black holes, pulsars and low-magnetic field neutron stars (Zezas et al., in preparation). White diamonds with black outlines and error bars indicate SMC sources with $\geq$3$\sigma$ detection in the full 4--25 keV \NuSTAR\ band. Black dotted lines show empirical boundaries between different compact-object types in color-color space, following \citet{Vulic2018}. We note the small error bars on the sources in our sample due to the large number of source counts. The lowest luminosity source in our 3$\sigma$ significance sample has $\sim$1500 net counts in the 4--25 keV band while the brightest sources have over 100,000 net counts. The error bars plotted represent 0.4 to 10\% errors in the 4--25 keV count rates. Note that source 1705 is a combination of two pulsars: previously confirmed SXP15.3 and the newly confirmed pulsar SXP305, which is presented in this paper. See \S \ref{new_pulsar} for more details.}
    \label{colorcolor}
\end{figure*}

\begin{figure*}
    \centering
    \includegraphics[width=0.9\textwidth,keepaspectratio]{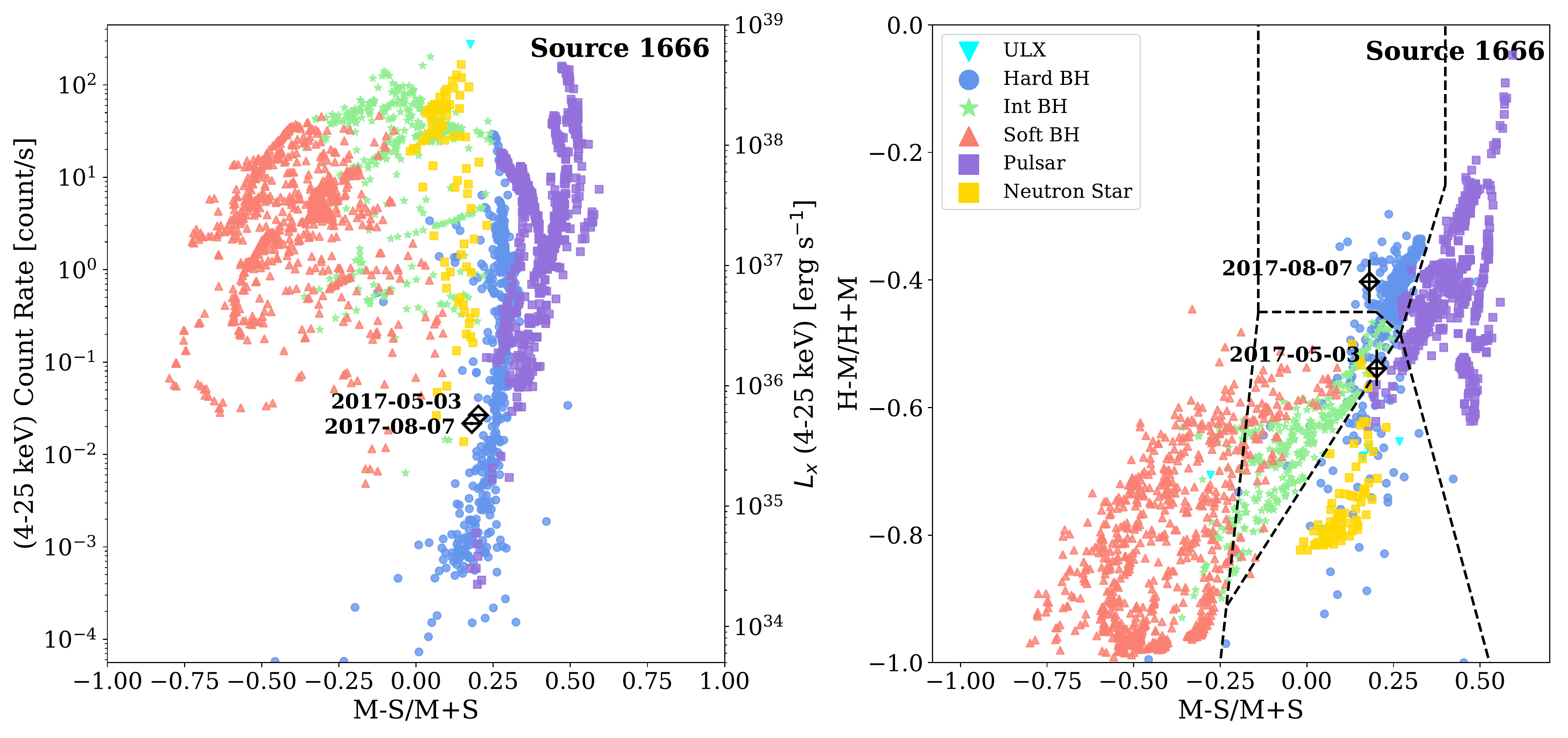}
    \caption{Hardness intensity and hardness ratio diagrams showing Source 1666 (source number indicated in upper right corner of each panel) at each observing epoch. The date of each observation is included as a label on the plot in yyyy-mm-dd format. For more information on background points, see caption of Figure \ref{colorcolor}. Source count rates and hardness ratios for each observation are listed in Table \ref{nustar_time}.}
    \label{var_1666}
\end{figure*}

\begin{figure*}
    \centering
    \includegraphics[width=0.9\textwidth,keepaspectratio]{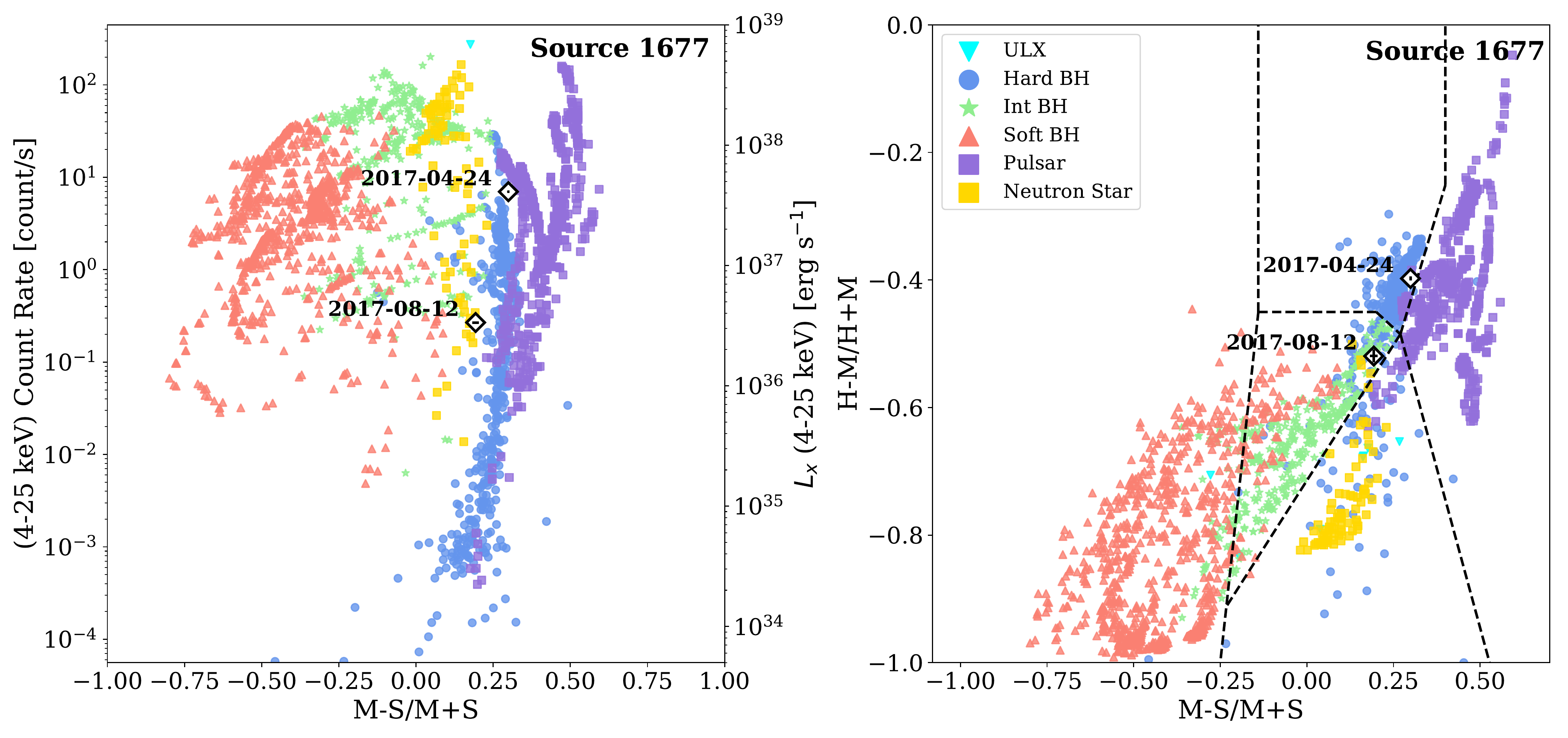}
    \caption{Hardness-intensity and hardness ratio diagrams for source 1677 at each observing epoch. See caption of Figure \ref{colorcolor} for more information on background points. Count rates and hardness ratios for this source during each observation are listed in Table \ref{nustar_time}. }
    \label{var_1677}
\end{figure*}

\begin{figure*}
    \centering
    \includegraphics[width=0.9\textwidth,keepaspectratio]{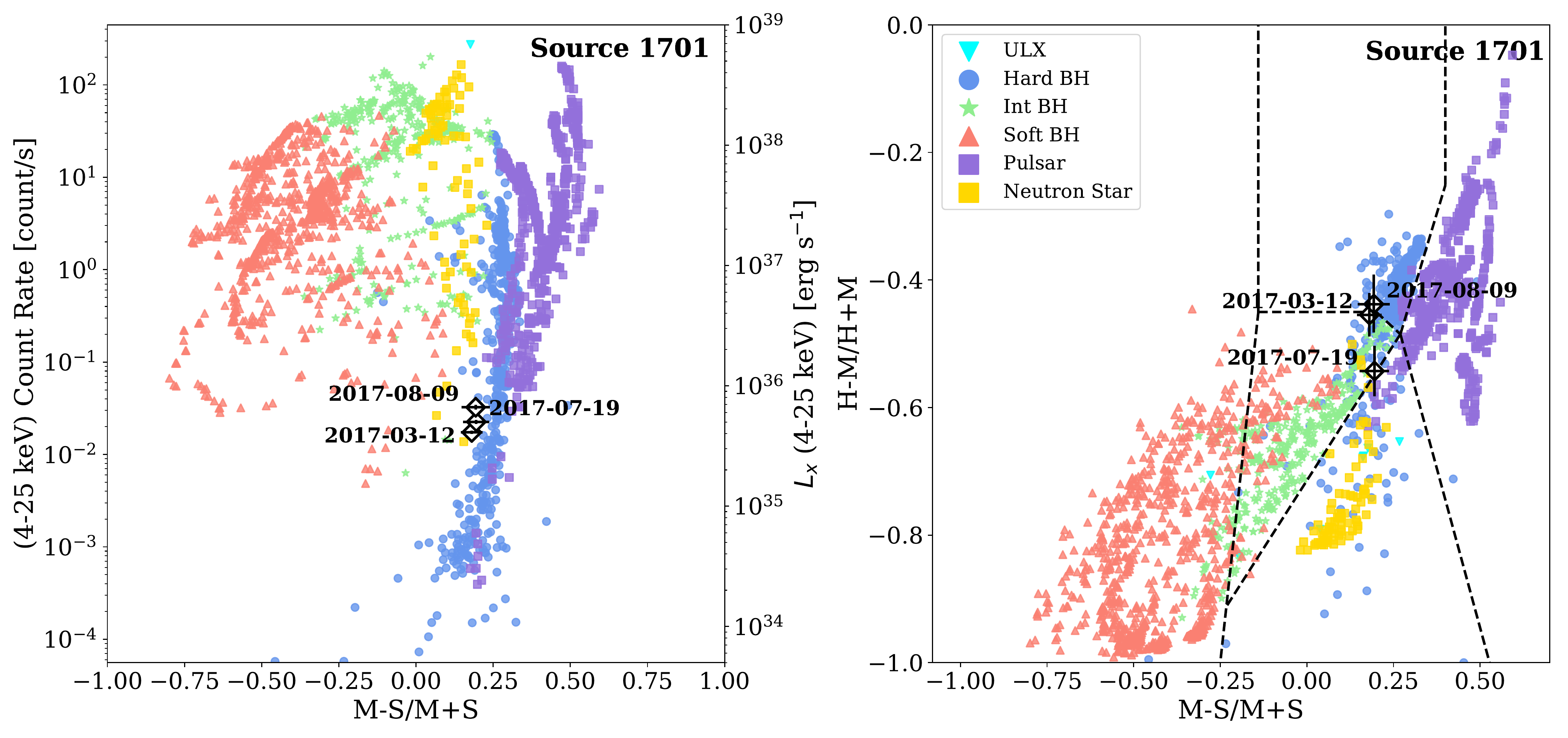}
    \caption{Hardness-intensity and hardness ratio diagrams for source 1701 at each observing epoch. See caption of Figure \ref{colorcolor} for more information on background points. Count rates and hardness ratios for this source during each observation are listed in Table \ref{nustar_time}. }
    \label{var_1701}
\end{figure*}

\begin{figure*}
    \centering
    \includegraphics[width=0.9\textwidth,keepaspectratio]{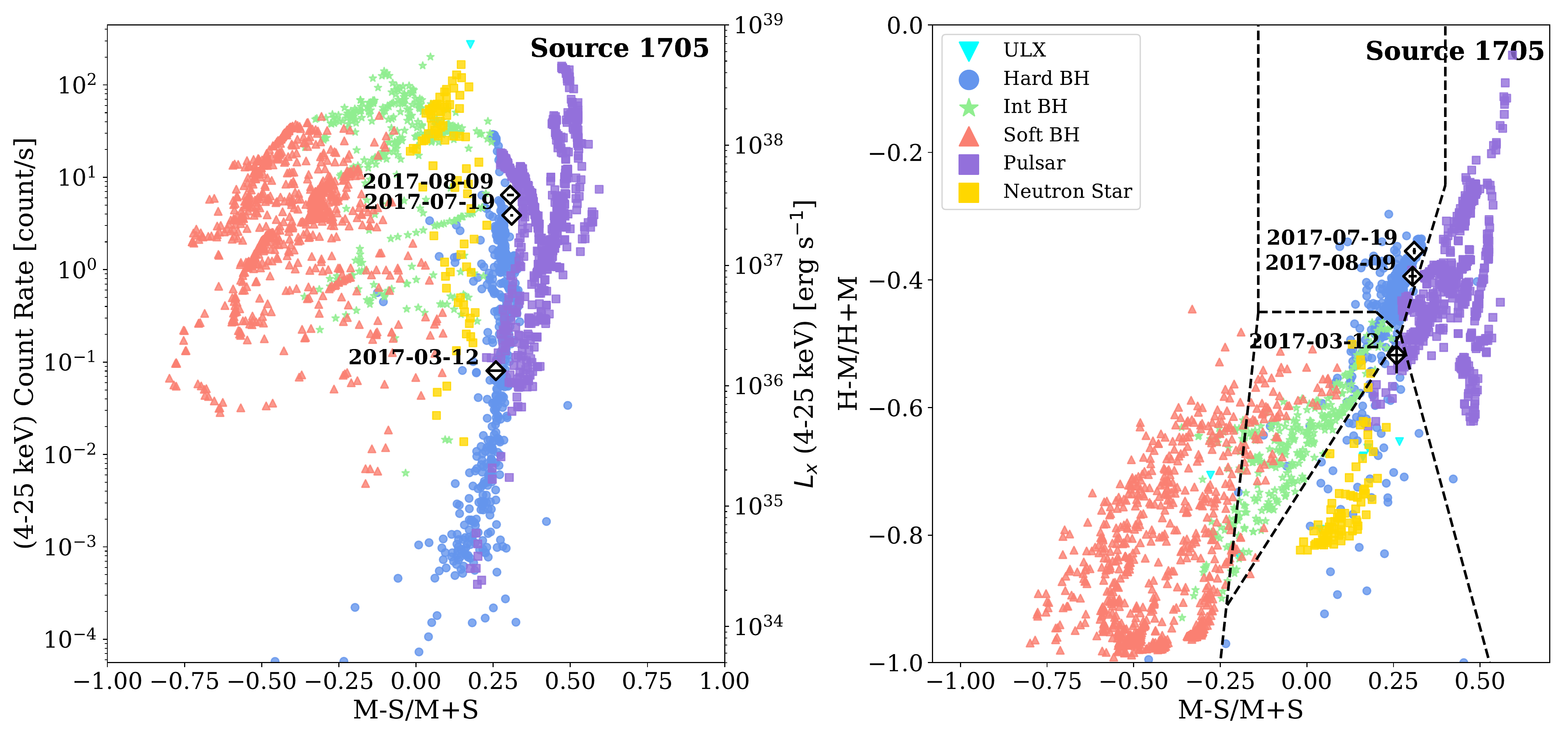}
    \caption{Hardness-intensity and hardness ratio diagrams for source 1705 at each observing epoch. See caption of Figure \ref{colorcolor} for more information on background points. Count rates and hardness ratios for this source during each observation are listed in Table \ref{nustar_time}.  We note that source 1705 is actually two sources - SXP 15.3 and the new pulsar presented in \S \ref{new_pulsar}, SXP305. SXP305 is active during the first observation of Field 2 (2017-03-12) and SXP15.3 is active during the second two observations (2017-07-19, 2017-08-09).}
    \label{var_1705}
\end{figure*}

\begin{figure*}
    \centering
    \includegraphics[width=0.9\textwidth,keepaspectratio]{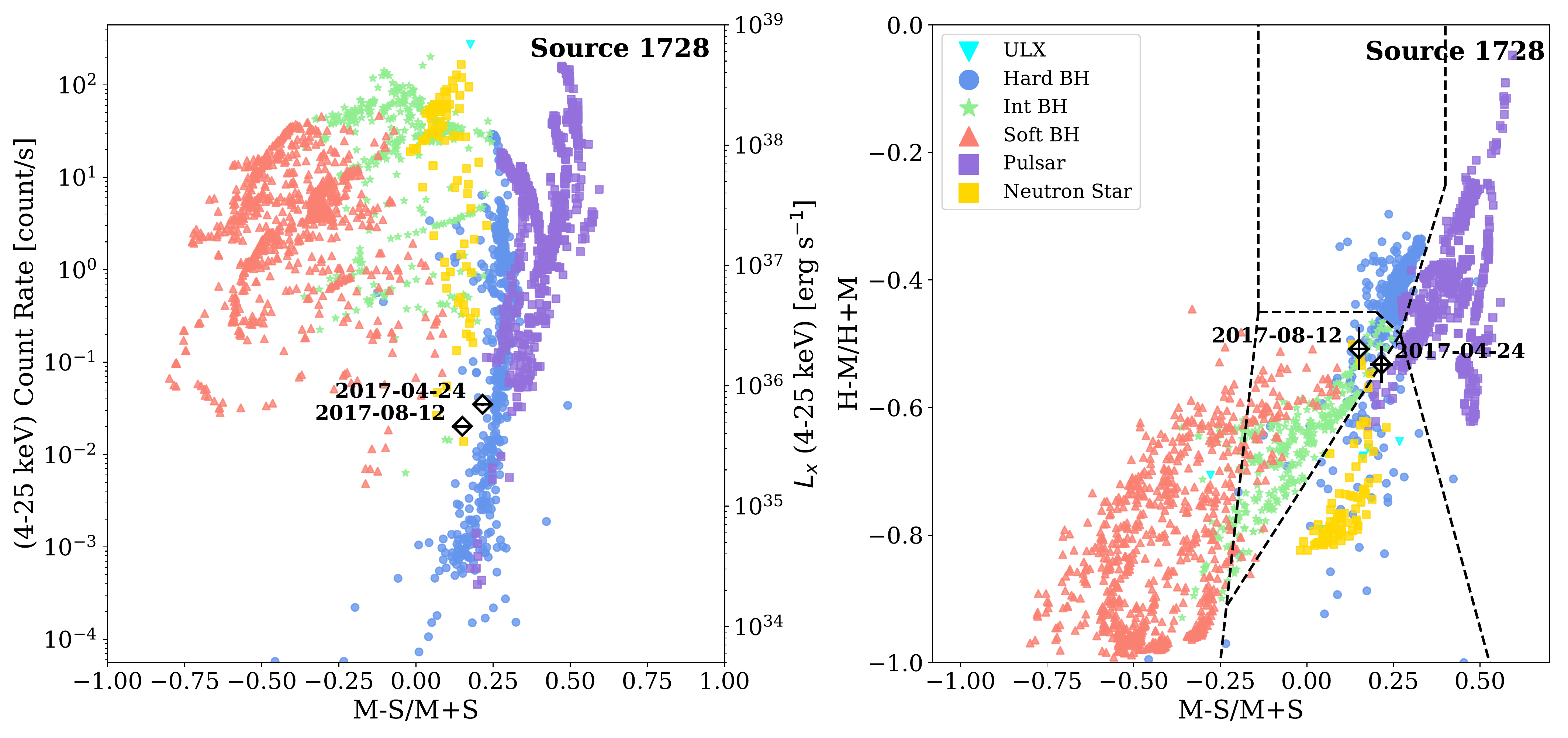}
    \caption{Hardness-intensity and hardness ratio diagrams for source 1728 at each observing epoch. See caption of Figure \ref{colorcolor} for more information on background points. Count rates and hardness ratios for this source during each observation are listed in Table \ref{nustar_time}. }
    \label{var_1728}
\end{figure*}

\begin{figure*}
    \centering
    \includegraphics[width=0.9\textwidth,keepaspectratio]{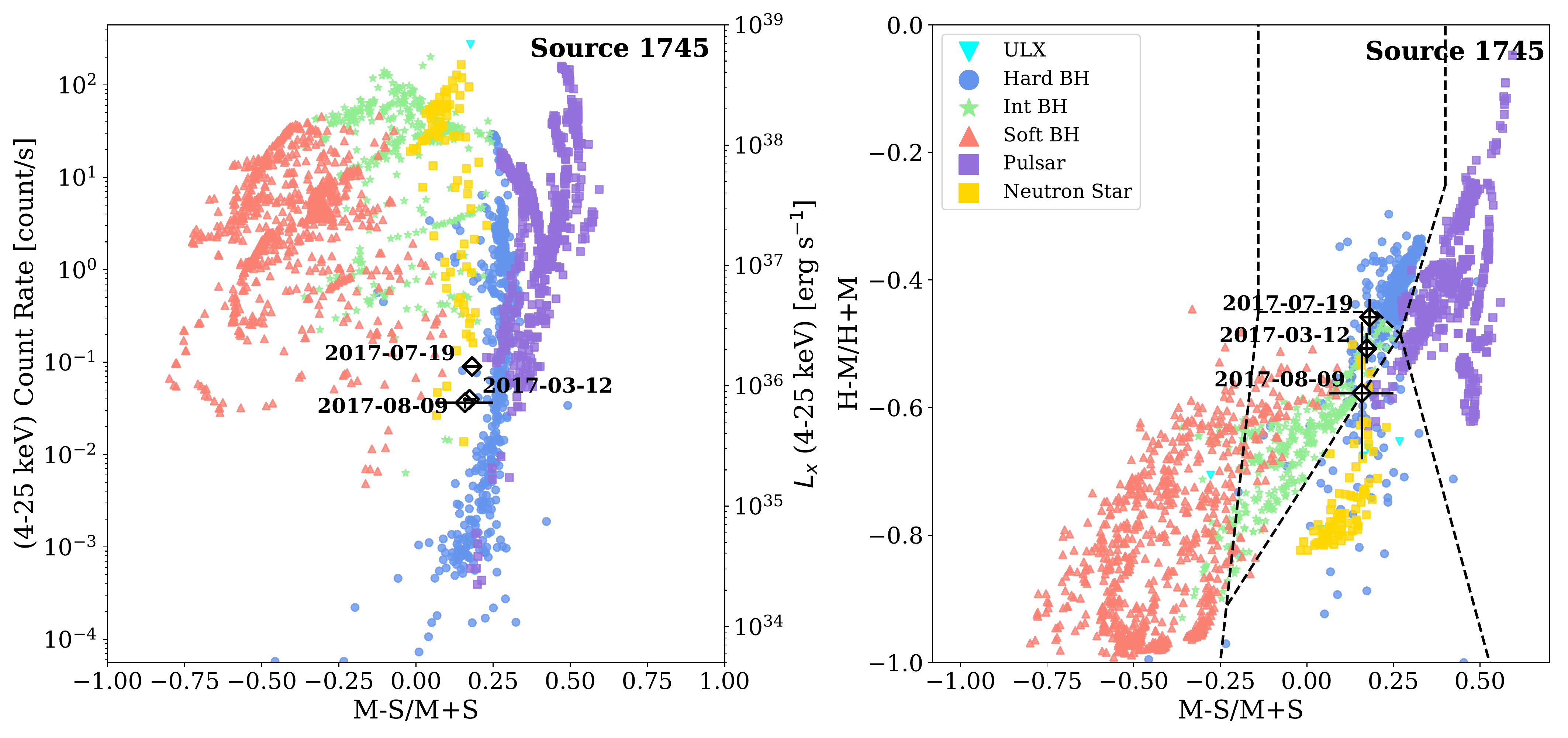}
    \caption{Hardness-intensity and hardness ratio diagrams for source 1731 at each observing epoch. See caption of Figure \ref{colorcolor} for more information on background points. Count rates and hardness ratios for this source during each observation are listed in Table \ref{nustar_time}. }
    \label{var_1731}
\end{figure*}

\begin{figure*}
    \centering
    \includegraphics[width=0.9\textwidth,keepaspectratio]{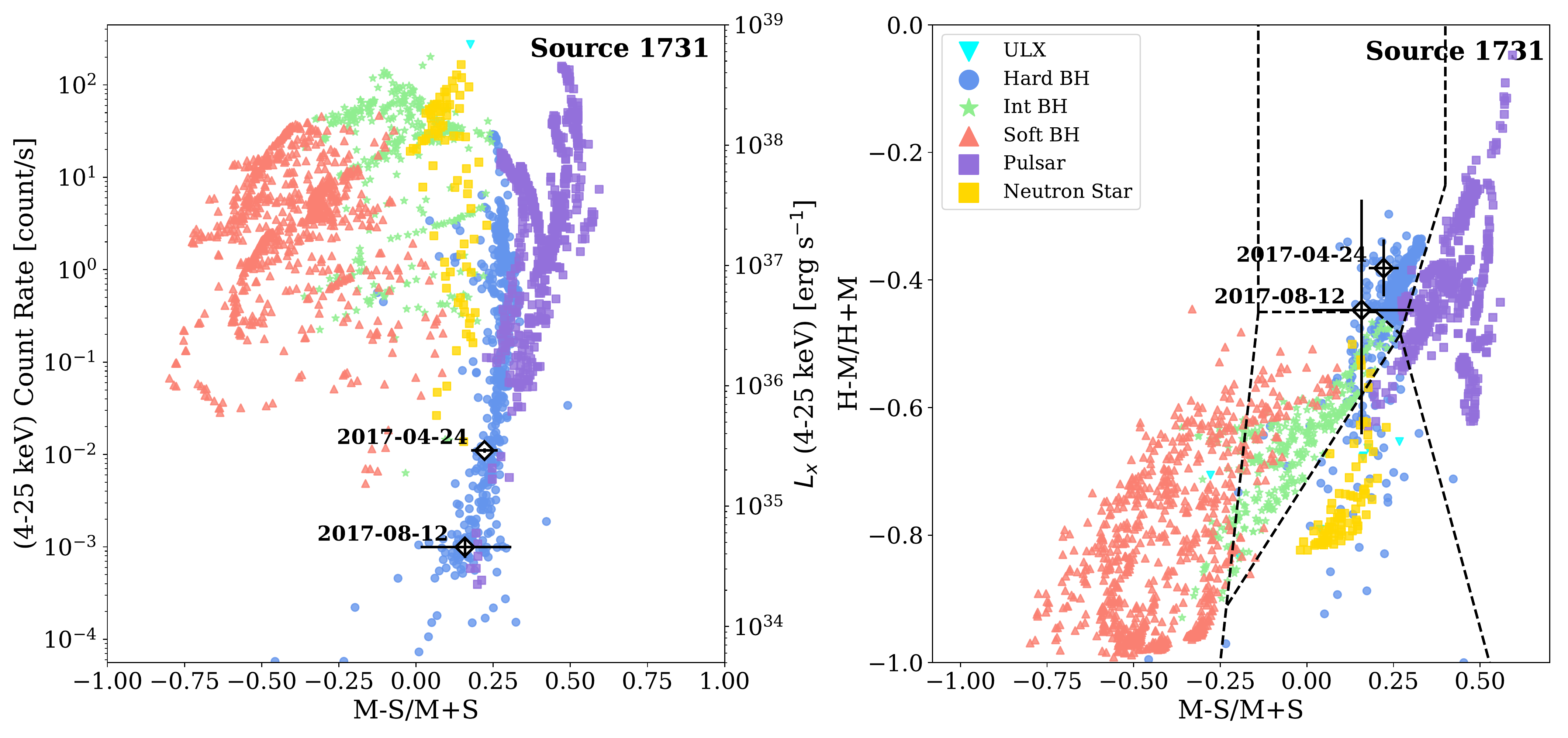}
    \caption{Hardness-intensity and hardness ratio diagrams for source 1745 at each observing epoch. See caption of Figure \ref{colorcolor} for more information on background points. Count rates and hardness ratios for this source during each observation are listed in Table \ref{nustar_time}. }
    \label{var_1745}
\end{figure*}

\begin{figure*}
    \centering
    \includegraphics[width=0.9\textwidth,keepaspectratio]{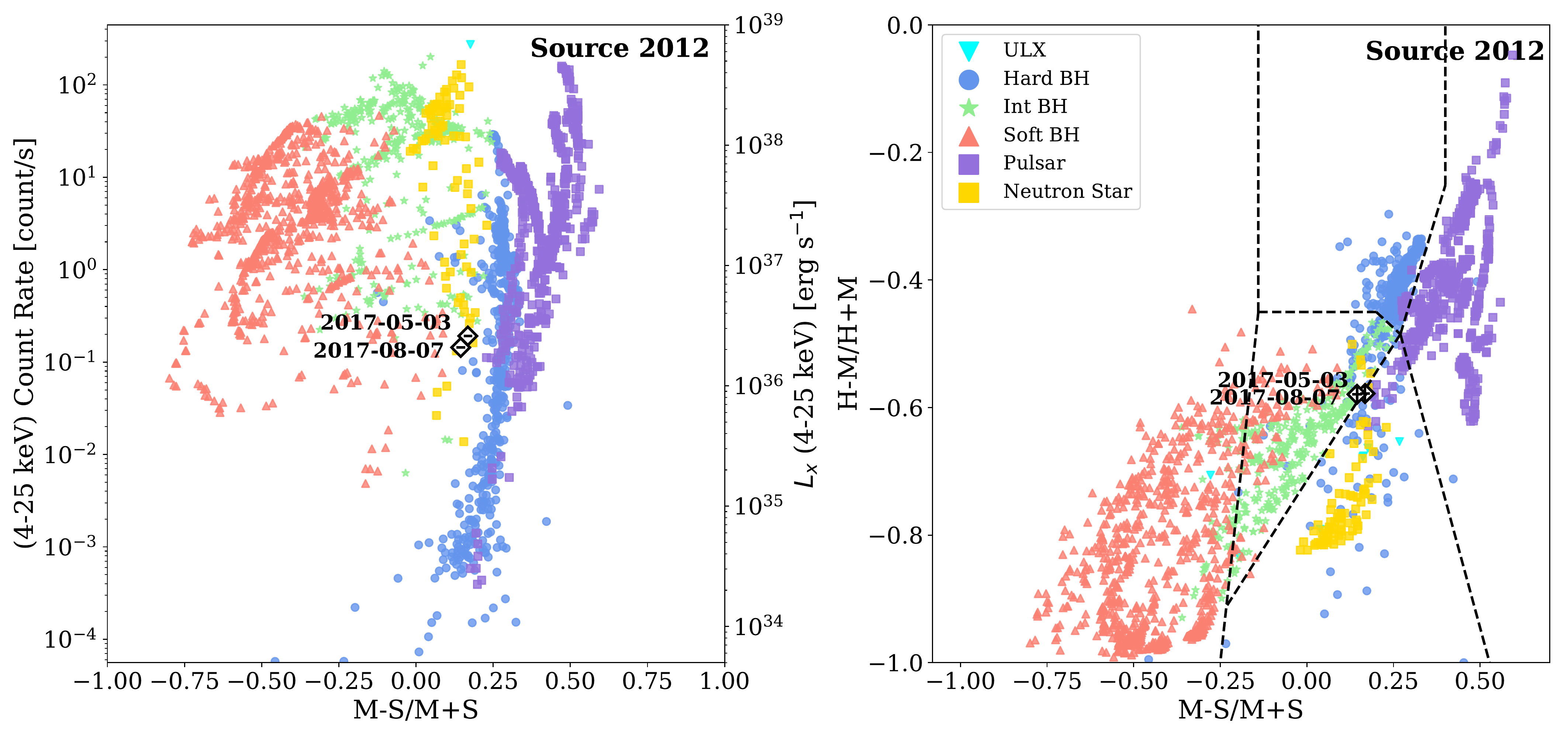}
    \caption{Hardness-intensity and hardness ratio diagrams for source 2012 at each observing epoch. See caption of Figure \ref{colorcolor} for more information on background points. Count rates and hardness ratios for this source during each observation are listed in Table \ref{nustar_time}. }
    \label{var_2012}
\end{figure*}

\begin{figure*}
    \centering
    \includegraphics[width=0.9\textwidth,keepaspectratio]{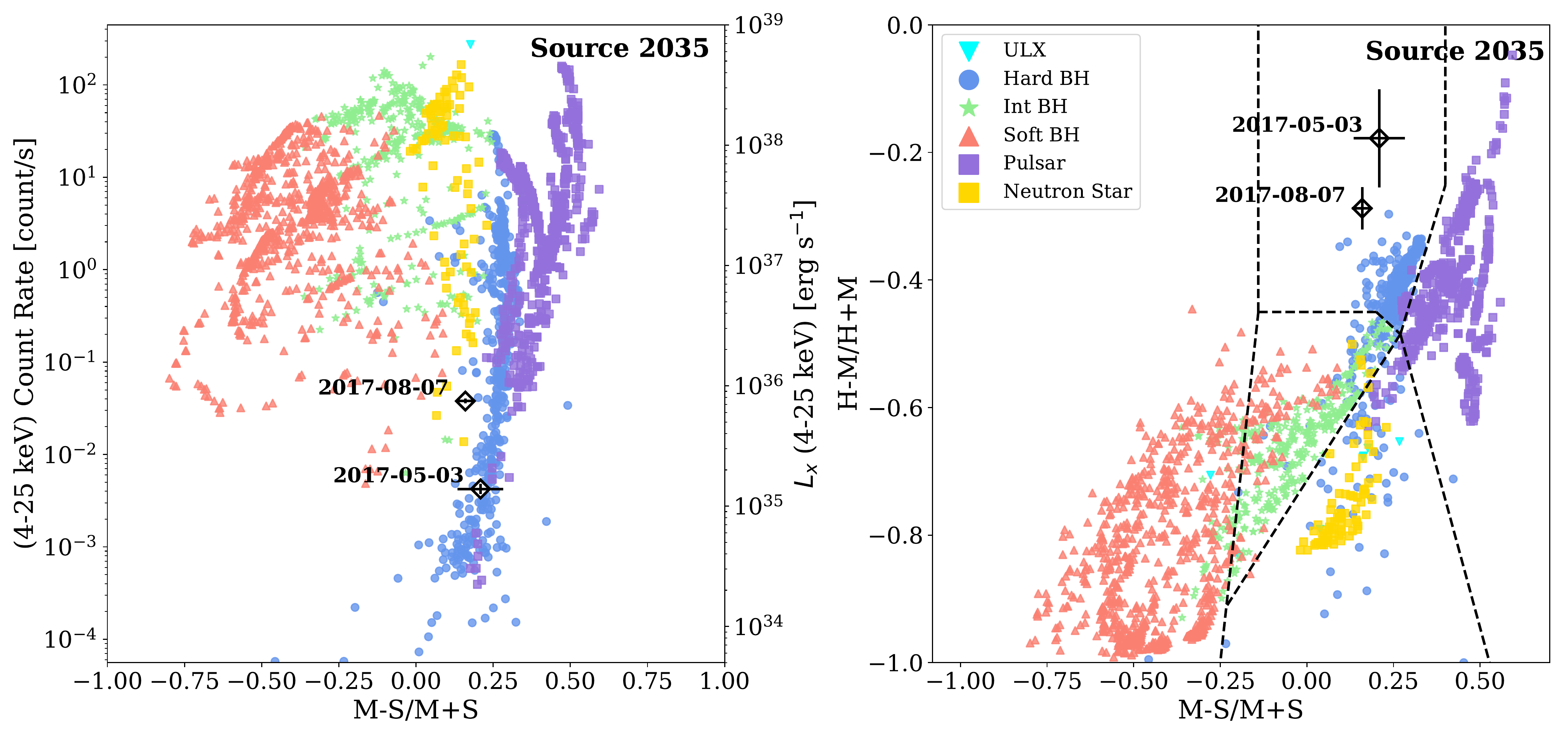}
    \caption{Hardness-intensity and hardness ratio diagrams for source 2035 at each observing epoch. See caption of Figure \ref{colorcolor} for more information on background points. Count rates and hardness ratios for this source during each observation are listed in Table \ref{nustar_time}. }
    \label{var_2035}
\end{figure*}

We classify XRBs in the SMC by comparing their X-ray luminosities and hardness ratios with those of Galactic XRBs with known compact-object types. The diagnostic diagrams that were used to classify each source are presented in Figures \ref{var_1666}--\ref{var_2035}, which plot the position of each source on the hardness-intensity and hardness ratio diagrams during each epoch of observation. For a more general overview of our sample, we also present a set of diagnostic diagrams where we plot count rate and hardness ratios for each source when all epochs of observation are combined in Figure \ref{colorcolor}. We note that due to variability between observations, we do not use the combined diagram (Figure \ref{colorcolor}) for our source classification.

The Galactic XRBs used in the diagnostic diagrams were observed with RXTE, not \NuSTAR, and their count rates were corrected for the different responses of the two missions. This method was developed by Zezas et al. (in preparation) and has been used previously to classify compact-objects in NGC 253 \citep{Wik2014}, M83 \citep{Yukita2016}, M31 \citep{Lazzarini2018}, Holmberg II, IC 342, M82, M81, NGC 4945, Holmberg IX, Circinus, NGC 1313, and NGC 5204 \citep{Vulic2018}. With better statistics, we can use the full spectra to gain even more information, separating black holes and neutron stars effectively \citep{Maccarone2016}.

Beyond classifying an XRB as having a black hole or neutron star primary, we can further classify black holes by accretion state (soft, intermediate, hard). The difference in spectrum can be used to infer changes in the dominant emission mechanism in each state. For black holes in the hard state, emission is dominated by a power law component from the optically thin region inside of and around the optically thick accretion disk. In the soft state, softer thermal blackbody emission from the optically thick disk dominates. The intermediate state is a shorter-lived transient state between the soft and hard states during which the luminosity remains fairly constant while the hardness ratio shifts. These differences in emission spectra allow hardness ratios, in combination with full band luminosities, to be used to discriminate between different black hole accretion states \citep[e.g.,][]{Remillard&McClintock}

Pulsars and Z-track NS XRBs are also included in the hardness-intensity diagnostic diagram. Low magnetic field neutron stars inhabit a narrow region of the hardness-intensity diagram, varying mostly in luminosity rather than X-ray color. Accreting pulsars generally exhibit harder X-ray spectra than even hard state black holes -- with a power law index of approximately 1 -- allowing for their separation from accreting black holes in hardness ratio parameter space \citep{White1983}. Low magnetic field neutron stars sources have softer X-ray spectra than pulsars \citep{Hasinger1989}.

The differences in hardness and luminosity in different black hole accretion states and neutron star types allows us to use these parameters to classify XRBs of unknown compact-object type. To create a diagnostic tool that can be used for \NuSTAR\ sources, A. Zezas et al. (2019, in preparation) completed spectral fitting for 6 BH-XRBs and 9 accreting pulsars using over 2500 RXTE PCA observations \citep{Sobolweska2009,Reig2011}. Different spectral models were applied to these spectra depending on their accretion state (i.e. the contribution of the thermal and power law components). These spectral models were then used to predict each source count rate in the $S$, $M$, $H$, and $F$ \NuSTAR\ bands. The 4--25 keV energy range used in our \NuSTAR\ observations falls within the energy range of the RXTE-PCA spectra, ensuring that the spectral models can adequately predict the \NuSTAR\ count rate in this energy range.

To classify the sources in the SMC, and thus determine the compact-object type, we examine their position on the diagnostic diagrams (Figures \ref{var_1666}--\ref{var_2035}), taking their error bars into account. For sources with error bars spanning multiple compact-object types, we list all possible compact-object types/states. All sources have two or three epochs of observation, which we plot separately on the diagrams to account for variability in both count rate and hardness ratios between observations. For sources with significant variability, we list source classifications consistent with all epochs of observation. We summarize our classifications in Table \ref{class_table}.

We note that because the field of view covered for this survey is much larger than for previous extragalactic \NuSTAR\ surveys, the rate of background AGN in our observations is likely to increase. We discuss one likely background AGN in our sample in \S \ref{agn}.

\section{Results \& Discussion}\label{results}
The deep \NuSTAR\ observations of 3 fields along the Small Magellanic Cloud Bar, resulted in a catalog of 10 sources with greater than 3$\sigma$ significance (4--25 keV) and 40 additional sources with upper limits on the count rate. Table \ref{class_table} provides our tentative classifications of the source compact-object types based on their hardness ratios and hard X-ray luminosities. We plot sources on the diagnostic diagram using their count rates and add the luminosity axis assuming a power law model with a photon index of 1.7 and the mean Galactic column density from all three fields (see \S \ref{intro}).

\subsection{Comparison with Archival \XMM\ Observations}\label{XMM_Comparison}
When performing PSF fitting to measure \NuSTAR\ count rates for sources in our observed fields, we used as priors the \Chandra\ source catalog from \citet{Antoniou2019} for initial source positions. We used input sources down to $\sim 5 \times 10^{33}$ erg s$^{-1}$ in the 0.5-8.0 keV \Chandra\ band, which corresponds to $\sim$ 15 times below our \NuSTAR\ detection limit, correcting for bandpass differences. Of the 50 source positions we input into our PSF fitting routine, 10 sources had measured count rates above the 3$\sigma$ detection limit, while 40 sources were non-detections and are presented as upper limits.

\begin{figure*}
    \centering
    \includegraphics[width=0.8\textwidth,keepaspectratio]{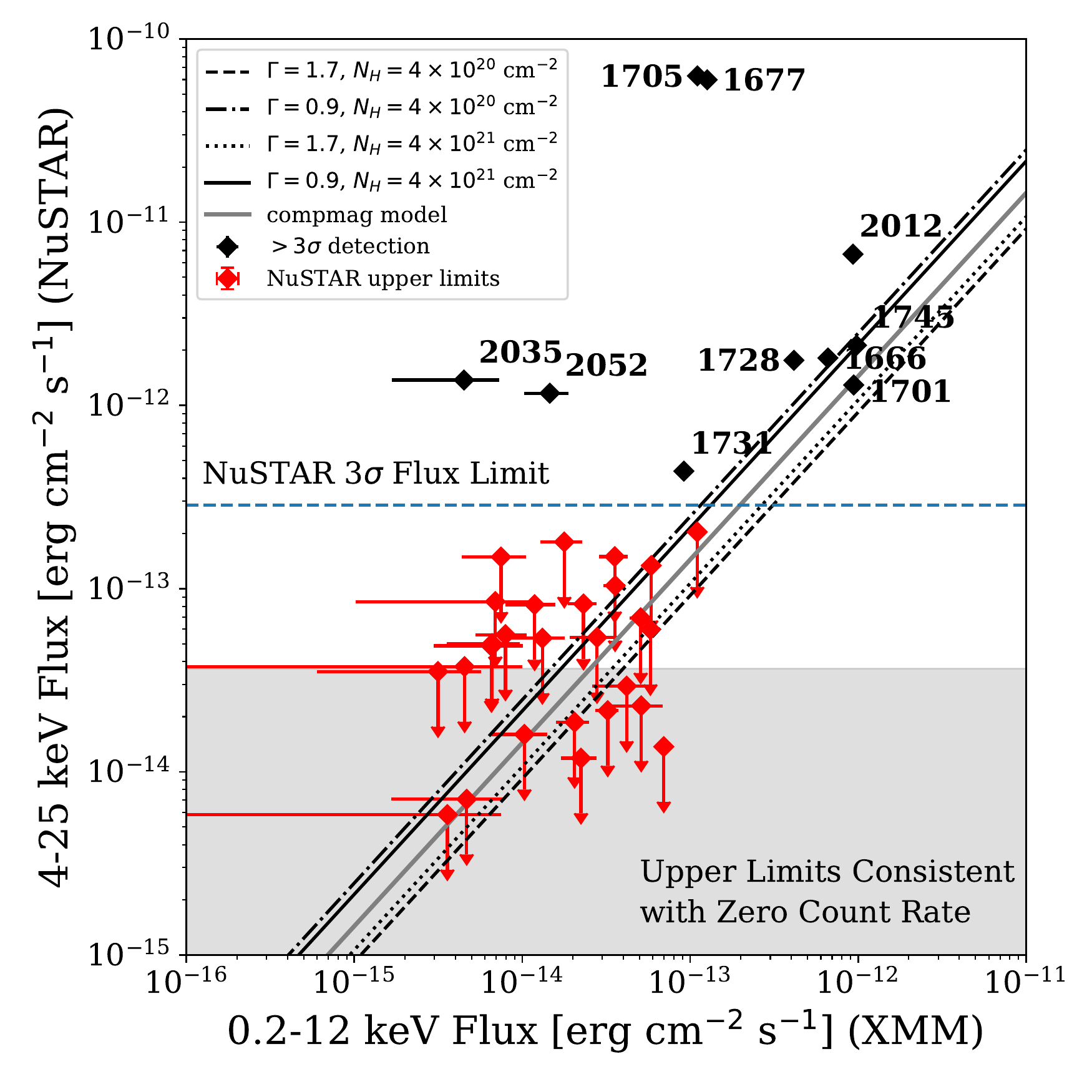}
    \caption{We compare measured 4--25 keV \NuSTAR\ source fluxes (combining all observations FMPA + FPMB data, omitting FPMB telescope for observations with stray light noted in Table \ref{obs_list}) and source flux upper limits with their 0.2-12 keV fluxes measured by \XMM\ \citep{Sturm2013}. Black points indicate sources with greater than 3$\sigma$ significance for their 4--25 keV count rates. Red points indicate the 1$\sigma$ upper limits for sources below the 3$\sigma$ detection limit. The horizontal blue line indicates the 3$\sigma$ flux limit for our observations. The gray shaded region indicates upper limits corresponding to zero measured count rate. We note that the spread in the upper limits of the red points corresponds to the expected scatter of low signal to noise measurements for these sources. The diagonal lines represent the relationship between 4--25 keV and 0.2-12 keV flux for various spectral models. The first four models in the legend assume a simple power law with the given photon index and Galactic column density. The fifth model represents a more physically motivated model for a low luminosity pulsar observed with \NuSTAR\ by \citet{Ballhausen2017}. For more details on the comparison between \NuSTAR\ and \XMM\ flux measurements, see \S \ref{XMM_Comparison}. Note that source 1705 is a combination of two pulsars: previously confirmed SXP15.3 and the newly confirmed pulsar SXP305, which is presented in this paper. See \S \ref{new_pulsar} for more details.}
    \label{xmm}
\end{figure*}

We investigated whether the sources that were not detected in our \NuSTAR\ observations would have expected 4--25 keV fluxes below our detection limit based on their independent flux measurements in other energy bands with another telescope. In order to determine whether we would expect non-detections for these sources, we cross matched our \NuSTAR\ source catalog with the \XMM\ survey of the SMC \citep{Sturm2013}. We positionally cross matched sources within 5$^{\prime \prime}$. We find 38 matches between the \citet{Sturm2013} catalog and our full \NuSTAR\ catalog, including sources with upper limits on \NuSTAR\ flux. In Figure \ref{xmm} we plot the 0.2-12 keV flux measured by \XMM\ \citep{Sturm2013} against our measured 4--25 keV flux (or upper limits on flux, where applicable). 

Black points in Figure \ref{xmm} are our 3$\sigma$ detections and sources plotted in red are 1 $\sigma$ upper limits on flux from our \NuSTAR\ observations. We plot a horizontal blue line that indicates the flux limit for a 3$\sigma$ detection. The gray region on our diagram indicates the upper limits on flux that are consistent with measurements of zero counts from a source. We perform PSF fitting at the location of all input \Chandra\ sources, so we obtain either a measured or a zero count rate detection for each source, with errors. The upper limits within the gray region show the upper errors on a zero count rate measurement. We note that sources within the gray region correspond to non-detections, while red sources outside of the gray region may potentially be detected at very low significance. The scatter in the upper limit values for the low signal to noise detections is expected scatter in these measurements.

We also include lines indicating the relationship between 0.2-12 keV flux and 4--25 keV flux for various spectral models by using \verb!XSPEC v. 12.10.0c!. The first four lines in the legend assume a simple power law model with a hard power law index (0.9) and softer power law index (1.7) and high Galactic absorption ($4\times10^{21}$ cm$^{-2}$) and low Galactic absorption ($6\times10^{20}$ cm$^{-2}$). The fifth line indicates the predicted 4--25 keV flux assuming the \verb!compmag! model in \verb!XSPEC!. The \verb!compmag! model is used by \citet{Ballhausen2017} to fit the 4--25 keV \NuSTAR\ spectrum of a low luminosity pulsar observed with \NuSTAR, A 0535+26. The model is cited as a more physical, rather than empirical, fit to a low luminosity pulsar spectrum. It includes cylindrical accretion onto a magnetized neutron star, including different velocity profiles and the second-order bulk Comptonization term in scattering calculations. We list the conversion factors that were used to generate the lines for each spectral model shown in Figure \ref{xmm} in Table \ref{conversion}. 

We expect non-detections for all red sources in Figure \ref{xmm} in the 4--25 keV \NuSTAR\ band because their 0.2-12 keV fluxes measured by \citet{Sturm2013} suggest that their 4--25 keV fluxes are below our detection limit for all spectral models. We note that several of our $>$3$\sigma$ sources have higher than expected 4--25 keV \NuSTAR\ fluxes, for all spectral models. This difference is likely due to source variability.

\subsection{Classifying Low Luminosity HMXBs}\label{low-lx pulsars}
Many of the hard X-ray sources we detected in this sample are spatially coincident with confirmed pulsars within 5$^{\prime \prime}$. In our classifications listed in Table \ref{class_table}, many of our sources have luminosities and hardness ratios consistent with multiple compact-object types, including accreting black hole primaries. Sources 1728, 1701, 1666, 2012, and 2035 are found in regions of the diagnostic diagrams (Figures \ref{var_1728}, \ref{var_1701}, \ref{var_1666}, \ref{var_2012}, \ref{var_2035}) consistent with multiple compact-object types - yet all of these XRBs are associated with known X-ray pulsars \citep{Haberl2016}. Hereafter, we refer to these sources as inconsistent pulsars, as relates to the \NuSTAR\ hardness-intensity diagram.

The SMC presents a unique opportunity to observe low-luminosity accreting pulsars. Given its proximity, we are attempting to classify XRBs in the SMC at lower luminosities than has been possible in previous extragalactic \NuSTAR\ studies. Previous work in M31 had a 4--25 keV luminosity limit of $\sim 3\times 10^{36}$ erg s$^{-1}$ \citep[Wik et al., in prep.;][]{Lazzarini2018} and more distant galaxies surveyed by \citet{Vulic2018} had 4--25 keV luminosity limits of at least $1\times10^{37}$ erg s$^{-1}$. 

The low-luminosity (L$_{4-25 keV}<1\times10^{37}$ erg s$^{-1}$) pulsars we observe tend to have softer $(M-S)/(M+S)$ hardness ratios than the Galactic accreting pulsars used to generate the diagram. There are many proposed differences between the HMXB populations at solar (Galactic) and sub-solar metallicity (SMC). Population synthesis studies predict that metallicity may affect HMXB populations in different ways. It has been suggested that at lower metallicity, HMXB populations may be more luminous due to hosting more massive compact objects \citep{Dray2006,Fragos2013}, have a higher fraction of Roche lobe overflow systems, and a different ratio of Be versus supergiant stellar companions \citep{Linden2010}. The underlying cause of these predicted differences between solar (Galactic) and sub-solar metallicity (SMC) HMXBs is the fact that lower-metallicity stars exhibit weaker radiatively driven winds.

Looking at the hardness ratio diagram (Figure \ref{colorcolor}, right panel), the $(H-M)/(H+M)$ colors of sources 1728, 1666, 1701, 2012, and 2035 all fall within roughly -0.3 and -0.6, a range that matches the Galactic pulsars used in the diagnostic diagram. The only difference appears in the $(M-S)/(M+S)$ hardness ratio. All of our inconsistent low-luminosity pulsars have a $(M-S)/(M+S)$ hardness ratio of $\sim 0.2$, while the Galactic accreting pulsars have slightly higher hardness ratios ranging from $\sim$0.3-0.5. As we describe in the next two sections, we cannot explain this inconsistency with X-ray variability. We also discuss in more detail the HMXB candidates that were detected above 3$\sigma$ significance that do not have confirmed pulsations in \S \ref{not_pulsars}. We have also thoroughly investigated whether source confusion or mis-matching might give rise to these differences. The overall good matching of \XMM, \NuSTAR, and \swift\ flux indicates we have identified the correct source matches.

\subsection{Highly Variable Sources}\label{variability}
Several of our high significance \NuSTAR\ sources have higher 4--25 keV fluxes than extrapolated from their 0.2-12 keV fluxes measured with \XMM\ by \citet{Sturm2013}, as shown in Figure \ref{xmm}. The \XMM\ and \NuSTAR\ observations were not simultaneous. \XMM\ observations were taken between 2000 and 2009 while \NuSTAR\ observations presented in this paper were taken in 2017. 

To confirm that these flux differences are due to variability, we compare our \NuSTAR\ measurements with the quasi-simultaneously measured 0.2-10 keV count rates from \swift. We positionally matched sources detected in each \NuSTAR\ observation with sources detected in each \swift\ observation within 10$^{\prime \prime}$. 

Nine of our ten 3$\sigma$ sources were detected in the quasi-simultaneous \swift\ observations. For more details on the \swift\ observations and data reduction, see \S \ref{swift_observations}. Only one of our 3$\sigma$ \NuSTAR\ sources was not detected by \swift, source 2052. Source 2052 is a likely background AGN (see \S \ref{agn}), so it is likely more luminous at the higher energy range probed by \NuSTAR\ (4--25 keV) than \swift\ (0.2-10 keV) due to photoelectric absorption.

\begin{figure*}
    \centering
    \includegraphics[width=0.9\textwidth,keepaspectratio]{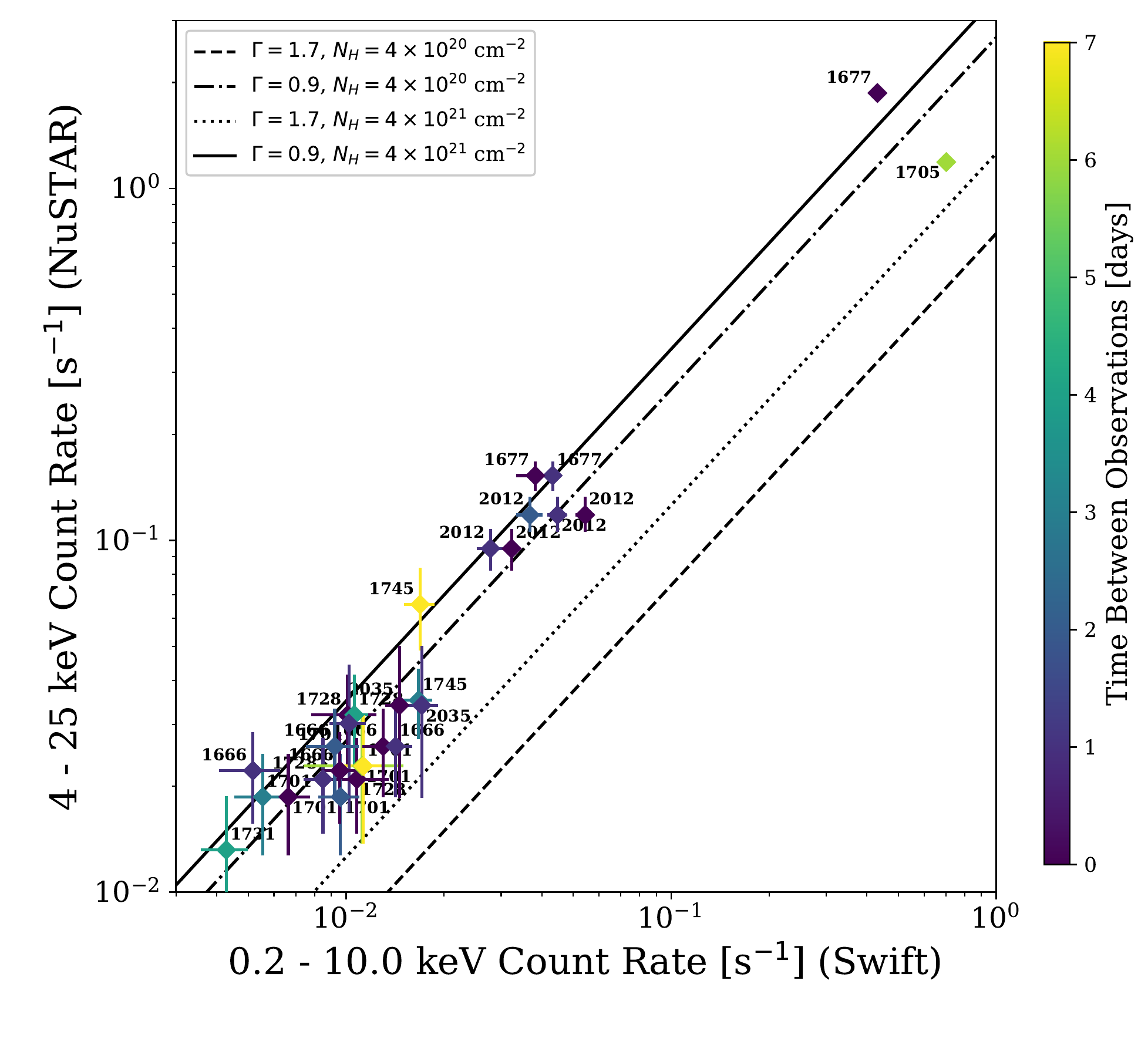}
    \caption{We compare measured 4--25 keV \NuSTAR\ source count rates (combining FMPA and FPMB data, omitting FPMB telescope for observations with stray light noted in Table \ref{obs_list}) and quasi-simultaneously measured 0.2-10 keV count rates from \swift. Each point represents a pair of roughly simultaneous observations, color coded by the time between observations in days. Sources are labeled with their Source IDs as listed in Table \ref{nustar_table}. Note that most sources are plotted more than once. Each point represents an individual observation. The count rates for each individual observation used to create this diagram are listed in Table \ref{nustar_time}. Lines on the plot represent the relationship between the \NuSTAR\ 4--25 keV count rate and the \swift\ 0.2-10 keV count rate assuming different power law spectral models, as described in the legend. Only sources that were detected by \swift\ are plotted here. Note that source 1705 is a combination of two pulsars: previously confirmed SXP15.3 and the newly confirmed pulsar SXP305, which is presented in this paper. See \S \ref{new_pulsar} for more details.}
    \label{swift_Comparison}
\end{figure*}

In Figure \ref{swift_Comparison}, each point indicates a pair of quasi-simultaneous measurements of a source's count rate by \NuSTAR\ (4--25 keV) and \swift\ (0.2-10 keV). Each point is labeled with the source number and is color coded by the time separation between the \NuSTAR\ and \swift\ observations. We list the count rates and hardness ratios measured for our 3$\sigma$ sources during each \NuSTAR\ observation in Table \ref{nustar_time}. We list the 0.2-10 keV count rates for each source detected by \swift\ in Table \ref{swift_data}. In Figure \ref{swift_Comparison} we also include lines that show the relationship between \NuSTAR\ 4--25 keV count rates and \swift\ 0.2-10 keV count rates for sources assuming various spectral models. Most of our sources fall along the two lines for $\Gamma=0.9$, expected for accreting pulsars. 

We note that some of the \swift\ and \NuSTAR\ observations that we compare in this figure were taken up to 7 days apart, with a median time separation of 1 day. This quasi-simultaneity makes it harder to directly compare the 0.2-10 keV count rates measured by \swift\ and the 4-25 keV count rates measured by \NuSTAR. Differences in the measured count rates could be due to source variability if the observations are not truly simultaneous.

Figures \ref{var_1666}--\ref{var_2035} show hardness-intensity and hardness ratio diagrams indicating the location of each source on the diagram during each observing epoch to investigate spectral shifts and variability between epochs. We include one diagram for all of our 3$\sigma$ sources except source 2052, which was not detected with high enough significance in each individual observation to produce good hardness ratio measurements.

In the following sections we describe several sources that demonstrated significant variability when we compare their \NuSTAR\ and \swift\ fluxes to those measured with \XMM\ in the \citet{Sturm2013} catalog.

\subsubsection{Source 1705 - Detection of two Be-XRBs}
Source 1705 falls along the boundary between the hard state black hole and pulsar loci on our diagnostic diagrams (Figure \ref{var_1705}) during each epoch of observation. Source 1705 is 0.5$^{\prime \prime}$ away from \CXOJ, listed as source 93 in \citet{Haberl2016}, and 7.5$^{\prime \prime}$ away from the known pulsar SXP15.3. Based on the results of our timing analysis (see \S  \ref{new_pulsar}) we suggest that during the first observation of source 1705 in March 2017 we detected flux predominantly from \CXOJ\ and during the next two observations in July and August 2017 we detected flux mostly from SXP15.3. 

SXP15.3 was found in outburst starting in July 2017 as part of the S-CUBED survey with \swift\ \citep{Kennea2018}. In November 2017 \citet{Ducci2017} observed the source in outburst. \citet{Maitra2018} observed the source in late 2017 with both \NuSTAR\ and \swift\ simultaneously and measured a 3--80 keV luminosity of $\sim 10^{38}$ erg s$^{-1}$. We obtained three \NuSTAR\ observations of this source in March, July, and August 2017, respectively (see Table \ref{nustar_time}). 

The outburst evolution of SXP15.3 found in the literature matches the flux variations seen in our observations. During the first epoch of our observations in March 2017, the 4--25 keV luminosity of source 1705 was $\sim 1\times10^{36}$ erg s$^{-1}$. By the second and third epochs in July and August 2017, source 1705's $4-25$ keV luminosity had increased to $3-4 \times 10^{37}$ erg s$^{-1}$. Our observations did not continue through the end of 2017, when SXP15.3 reached the peak of its outburst.

We note that \CXOJ, located within 0.5$^{\prime \prime}$ of the \NuSTAR\ position, was in outburst during the first epoch of our observations in March 2017. \CXOJ\ is a high confidence Be-XRB with a typical X-ray spectrum and an early-type optical counterpart \citep{Haberl2016}.

Our timing analysis of source 1705 revealed interesting results, depending on the epoch analyzed. During the first observation (obsid 50311002002) taken in March 2017, we detected pulsations with a period of $\sim$305 seconds. We suggest that during this first observation, what we were observing is associated with \CXOJ\ and that we are able to confirm it as Be-XRB pulsar. During the second and third observation epochs (obsids 50311002004, 50311002006) in July and August 2017 we detected a period of 15.3 seconds, which matches SXP15.3. We can not exclude that we were detecting flux from both Be-XRBs, \CXOJ\ and SXP15.3, with the flux from SXP15.3 dominating during the second two epochs when that source was known to be in outburst. 

\subsubsection{Source 1677 - Detection of SXP59.0}
Source 1677 is associated with known pulsar SXP59.0 (RX J0054.9$-$7226), with noted X-ray variability in the literature \citep{Haberl2016}. We measured a pulse period of 58.8 seconds using an averaged power spectrum (see Section \ref{timing} for more details on pulse fitting).

Source 1677 has a 4--25 keV flux roughly two orders of magnitude greater than would be expected given its 0.2-12 keV flux measured with \XMM, as shown in Figure \ref{xmm}. Its 4--25 keV luminosity also decreased by roughly one order of magnitude between our two epochs of observation, April and August 2017. 

Source 1677 lies in regions of the diagnostic diagram that overlap with pulsars, hard state black holes, intermediate state black holes, and non-magnetized neutron stars. Between the first observing epoch in April 2017 and August 2017, source 1677 became less luminous, and its spectrum became softer in both the $(M-S)/(M+S)$ and $(H-M)/(H+M)$ colors.

Source 1677 matches to source 63 in \citet{Sturm2013} within 0.8$^{\prime \prime}$. \citet{Sturm2013} note that this source demonstrates significant short-term variability in the 0.2-12 keV band. \citet{Haberl2016} cite that SXP59.0 has a ratio of 840 between its maximum and minimum X-ray flux presented in the literature. This extreme variability would account for the excess flux we see in our \NuSTAR\ observations compared to the \XMM\ observations from \citet{Sturm2013}. 

Source 1677 was also detected in two \swift\ observations within one day of the \NuSTAR\ observations. Both 0.2-10 keV count rate measurements by \swift\ agree with the 4--25 keV count rate measured simultaneously by \NuSTAR\ assuming a hard ($\Gamma \sim 0.9$) power law model. \citet{Kennea2017,SCUBED} discovered that SXP59.0 was in outburst in April 2017 with \swift\ observations, part of the S-CUBED survey. The 4--25 keV luminosity we measure with \NuSTAR\ also shows this source in outburst(4--25 keV $L_{X} \sim 10^{38}$ erg s$^{-1}$).

\subsubsection{Source 2035 - Detection of SXP46.6}\label{2035_variability}
Source 2035 is associated with known pulsar SXP46.6 (XTE J0053$-$724) noted in \citet{Haberl2016}. It also presents a higher measured 4--25 keV flux with \NuSTAR\ than we would expect given its 0.2-12 keV flux measured with \XMM\ by roughly a factor of 100.  

Source 2035 matches to source 1828 in \citet{Sturm2013} within 1.5$^{\prime \prime}$. \citet{Sturm2013} do not note this source as having significant short-term X-ray variability. However, \citet{Haberl2016} give a ratio between the maximum and minimum 0.2-10 keV flux in the literature for this source of 1300, suggesting that it is highly variable.

The system had a luminosity of a few $\times 10^{35}$ erg s$^{-1}$ during the first observation (obsid 50311003002) in May 2017 but reached a luminosity of close to $10^{36}$ erg s$^{-1}$ in August 2017.
We were able to confirm pulsations with a pulse period of 45.98 seconds only in the second \NuSTAR\ observation.
During the first observation a periodic signal of $\sim$58.8 s was derived from the extracted event files. We interpret that this signal is due to contamination from the nearby pulsar SXP\,59 that is only $\sim$4.5\arcmin away from the center of the extraction region.
We note that SXP46.6 did not have detected pulsations in the \Chandra\ X-ray Visionary Program survey \citep{Hong2017} when it was observed in 2006.

Source 2035 was detected in our \swift\ observations (see Figure \ref{swift_Comparison}) taken within one day of the \NuSTAR\ observations. The \swift\ flux measured for this source agrees with our measured \NuSTAR\ flux assuming a power law model with a photon index of 0.9. Given that both its soft X-ray flux measured with \swift\ and hard X-ray flux measured with \NuSTAR\ are higher than would be expected from the soft X-ray flux measured with \XMM\ in \citet{Sturm2013}, this highly variable source was likely caught in an outburst during the \NuSTAR\ observations.

Source 2035's position on the diagnostic diagrams (Figure \ref{var_2035}) changes between its two epochs of observation. In May 2017 its position on the hardness-intensity diagram straddles the loci associated with pulsars and hard state black holes, in a low-luminosity region of the diagram with few points associated with pulsars. In August 2017 its luminosity was higher and it moved to a region consistent with hard state black holes. Its position in the hardness ratio diagrams favors a hard state black hole classification but does not rule out a pulsar classification, particularly in the hardness-intensity diagram.

\subsubsection{Source 2052 - Likely AGN}\label{agn}
Source 2052 is likely a background AGN. It matches within 1.7$^{\prime \prime}$ to source 661 in \citet{Sturm2013} where it was classified as a likely AGN. \citet{Sturm2013} classify it as a likely background AGN because of its hard X-ray colors and the ratio of its X-ray to optical flux ($log(f_{X}/f_{o})$ is greater than -1, typical for an AGN \citep{Maccacaro1988}. However, this source does not appear in the \citet{Sturm2013b} catalog of background AGN in the SMC, which was based on identifications with radio sources. This source also does not appear in the catalog of newly identified AGN behind the SMC \citep{Maitra2019}, which on the other hand was X-ray/NIR selected. 

Source 2052 is roughly 100 times brighter in the 4--25 keV bandpass than its 0.2-12 keV flux measured by \XMM\ would indicate, assuming a power law model with a photon index of 1.7. Source 2052 was not detected in our quasi-simultaneous \swift\ observations. This is likely due to photoelectic absorption which would preferentially affect softer (E$<10$ keV) X-ray photons detected by \swift\ over harder (E$>10$ keV) X-ray photons detected by \NuSTAR.

\subsection{Sources without Confirmed Pulsations}\label{not_pulsars}

Of the 10 sources we attempted to classify 7 are confirmed pulsars, one is a likely background AGN (source 2052, see \S \ref{agn}), and two are previously identified HMXBs that do not have observations of pulsations in the literature. In the following sections we describe our observations of these two HMXBs.

\subsubsection{Source 1731 - HMXB}
Source 1731 is located within $\sim 1^{\prime \prime}$ of source 117 in the Haberl catalog \citep{Haberl2016}; XMMU J005618.8$-$722802. XMMU J005618.8$-$722802 was observed by \citet{Sturm2013} where it is identified as a HMXB candidate. \citet{Haberl2016} note that the source has measured Balmer (H$\alpha$) emission from its specturm. \citet{Shtykovskiy2005} first observed this source with \XMM\ and noted that it lies inside of the star cluster NGC 330.

The position of source 1731 changes on the diagnostic hardness-intensity and hardness ratio diagrams between epochs of observation (Figure \ref{var_1731}). During the first observation (April 2017) source 1731 occupies a region of the hardness-intensity diagram that has significant overlap between the hard state black hole locus and the pulsar locus. Its $(M-S)/(M+S)$ hardness ratio is softer than the pulsar locus of the diagram. During its second observation in August 2017, source 1731's luminosity is almost 10 times lower and its position on both the hardness-intensity and hardness ratio diagrams is consistent with pulsars and hard state black holes, within errors. We did not observe pulsations for source 1731 using an averaged power spectrum.

\subsubsection{Source 1745 - Low luminosity Be-XRB}
Source 1745 [source 84 in \citet{Haberl2016}] lies in the overlapping region between hard and intermediate state accreting black holes and accreting pulsars on the diagnostic diagrams. Thus far source 1745 has been identified as a Be-XRB \citep{MA93,Haberl2000b,Antoniou2009,Haberl2016} with a clear emission-line star as its optical counterpart. Its X-ray spectrum is typical of an X-ray binary with $\Gamma < 1.3$ \citep{Haberl2016}. This source does not have published variability information and does not have a detected pulse period in the literature. We did not observe pulsations for source 1745 using an averaged power spectrum.

\subsection{Pulsars}
\subsubsection{Source 1728 - Detection of SXP 645}
Source 1728 is associated with known pulsar SXP 645 (XMMU J005535.2$-$722906). We measured a pulse period of 625.0 seconds using an averaged power spectrum. This measured period is slightly shorter than the value from the literature, 645 seconds \citet{Haberl2008}. Source 1728 is difficult to classify with our diagnostic diagrams because it lies in a region that overlaps between hard state black holes and pulsars. It matches within 1.13$^{\prime \prime}$ to source 55 in \citet{Haberl2016}, where it is listed as a Be-XRB with an accreting pulsar primary.

\subsubsection{Source 1666 - Detection of SXP 138}
Source 1666 is associated with known pulsar SXP 138 (CXOU J005323.8$-$72271). We measured a pulse period of 138.9 seconds using an averaged power spectrum. This measured period agrees with the 138.04 $\pm$ 0.61 second period observed by \citet{Edge2004}. On our diagnostic diagrams, source 1666 lies in a region with overlap between hard state black holes and pulsars. It matches to source 30 in the \citet{Haberl2016} catalog within 0.6$^{\prime \prime}$, where it is identified as a Be-XRB with measured X-ray pulsations. Its companion star is a Be star \citep{Coe2005,Zaritsky&Harris2004}.

\subsubsection{Source 1701 - Detection of SXP323}
Source 1701 is associated with known pulsar SXP323 (RX J0050.8$-$7316). We measured a pulse period of 312.5 seconds using an averaged power spectrum. This measured period is slightly shorter than the published period of 323 seconds \citep{Imanishi1999}. Similarly to the other pulsars in our sample with $L_{4-25 keV}<1\times10^{37}$ erg s$^{-1}$, it lies in a region of our diagnostic diagram with overlap between accreting pulsars and hard state black holes. Source 1701 matches within 0.5$^{\prime \prime}$ to source 47 in \citet{Haberl2016}, which is identified as RX J0050.8$-$7316 (SXP323).

\subsubsection{Source 2012 - Detection of SXP7.77}
Source 2012 is associated with the known pulsar SMC X-3 (SXP 7.77). We measured a pulse period of 7.76 seconds using an averaged power spectrum, which agrees with the published period of 7.77 seconds \citep{Edge2004_X3}. On our diagnostic diagrams, source 2012 lies at the intersection of hard state black hole, pulsar, and intermediate state black holes on the hardness-intensity diagram. On the hardness ratio diagram, source 2012 lies at the soft (lower-left) corner of the pulsar locus. 

Source 2012 matches to SMC X-3, listed as source 9 in \citet{Haberl2016}, within 0.44$^{\prime \prime}$. SMC X-3 is a well documented accreting pulsar \citep{Li1977} with a Be optical counterpart \citep{Evans2004}. SMC X-3 was observed in a likely Type II outburst starting in 2016 and ending in February 2017, although it continued to be detected after the end of its outburst \citep{SMCX-3_first,SMCX-3_second,SCUBED,Koliopanos2018}. SMC X-3 has a well measured orbital period ($\sim45$ days) and measured X-ray variability that is consistent with Type I outbursts that peak at the XRB's orbital periastron \citep{Townsend2017}. We note that its hardness ratios and luminosity did not vary significantly between our two observations with \NuSTAR\ in May and August 2017 (see Fig. \ref{var_2012}).

\subsection{Detection of Pulsations from \CXOJ\ and its likely orbital period}\label{new_pulsar}

\begin{figure}
    \centering
    \includegraphics[width=0.5\textwidth,keepaspectratio]{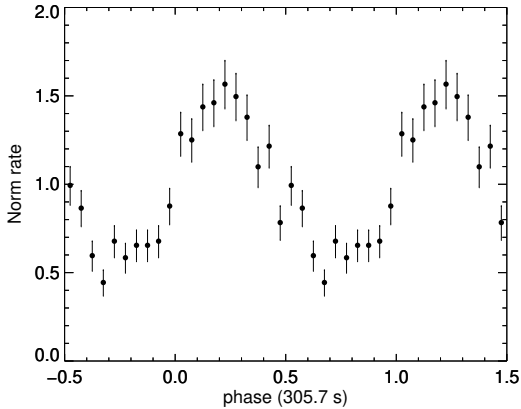}
    \caption{Pulse profile of the newly discovered pulsar SXP 305. The profile was fit for the full 4-25 keV \textit{NuSTAR} band and is background subtracted.}
    \label{fig:sxp304}
\end{figure}    

The small angular distance of \CXOJ\ to SXP\,15.3 led to initial confusion about the correct optical counterpart for SXP\,15.3 \citep[see the discussion on SXP\,15.3 in][]{Schurch2011}. Only after the detection of both X-ray sources in a Chandra observation \citep{Laycock2010}, it became clear that there are two Be-XRBs only 7.5\arcsec\ apart. A Swift observation nearly simultaneous to the first \NuSTAR\ observation shows that \CXOJ\ was active during the March 2017 observation. The pulse profile for our NuSTAR observations of CXO J005215.4-731915 is shown in Figure \ref{fig:sxp304}.

The optical counterpart of \CXOJ\ was observed by the Optical Gravitational Lensing Experiment (OGLE), which started observations in 1992 \citep{1992AcA....42..253U}. The star was monitored  during phases II (smc\_sc6.99991), III (smc100.1.43700) and IV (smc720.26.531) until today \citep[for OGLE-IV see][]{2015AcA....65....1U}. Observations are performed with the 1.3~m Warsaw telescope at Las Campanas Observatory, Chile. Images are taken in the V and I filter pass-bands and photometric magnitudes are calibrated to the standard VI system.

\begin{figure}
    \centering
    \includegraphics[width=0.45\textwidth]{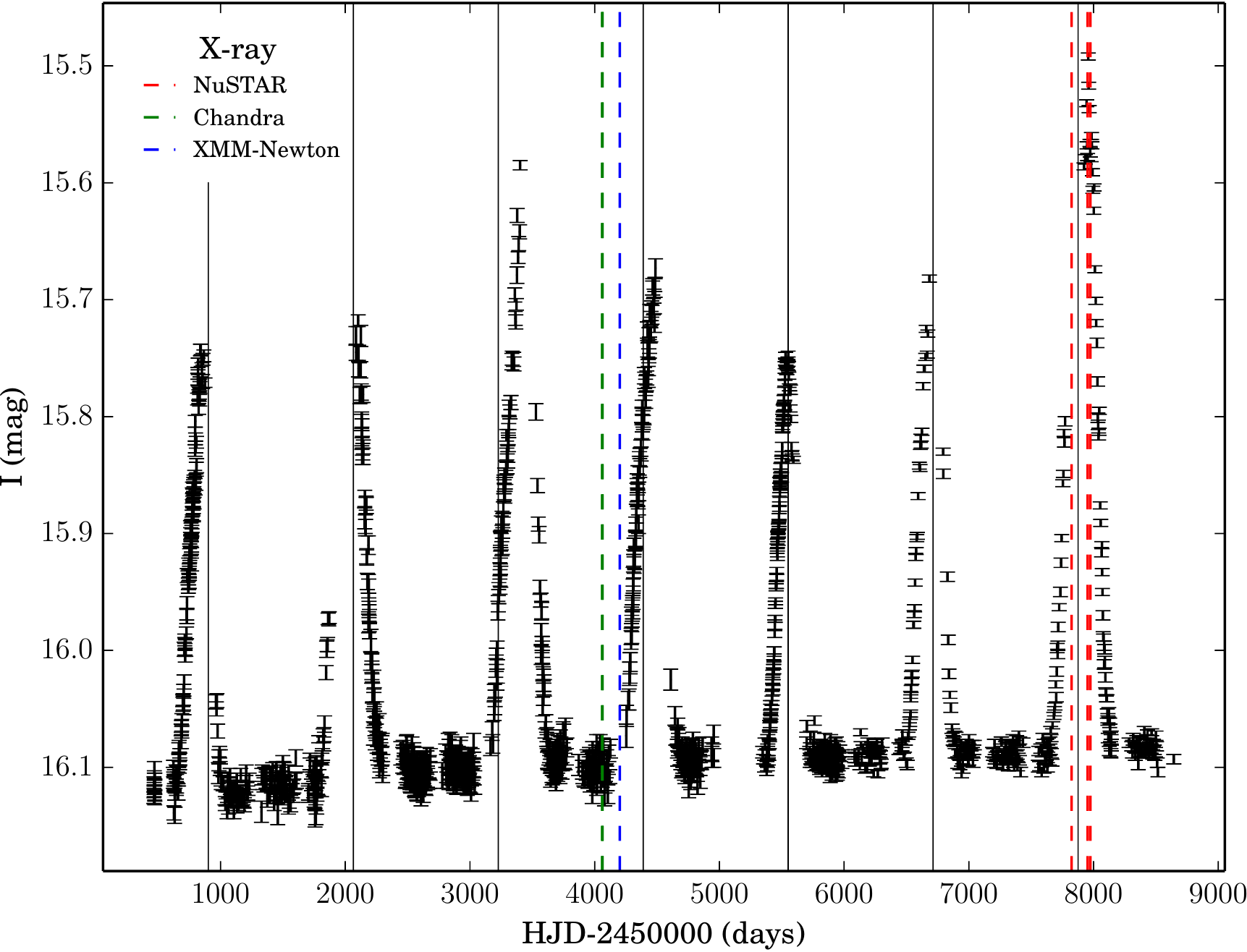}
    \caption{OGLE I-band light curve of the optical counterpart of \CXOJ. The colored vertical dashed lines mark the times of X-ray observations and the full black lines are separated by 1163 days.}
    \label{OGLE_LC}
\end{figure}

The OGLE II and III I-band light curve of the optical counterpart of \CXOJ\ was presented by \citet{Schurch2011} revealing regular outbursts by up to 0.5 mag. Figure~\ref{OGLE_LC} shows an updated light curve including the OGLE IV data. Seven outbursts are now recorded over more than 22 years. A Lomb-Scargle periodogram of the full light curve reveals a broad peak around $\sim$1163 days. A grid of intervals with this period (indicated by thin vertical lines in Fig.~\ref{OGLE_LC}) anchored on the peak of the fifth outburst (the only one with a fully covered peak) shows that the outbursts do not occur strictly periodic. A period of more than 1000 days is very long for the orbital period of a Be-XRB and \citet{Schurch2011} proposed that the outbursts are caused by changes in the structure and size of the circum-stellar disk. After detrending the OGLE light curve they suggest an orbital period of 21.68 days, detected in their Lomb-Scargle periodogram.

\begin{figure}
    \centering
    \includegraphics[width=0.45\textwidth]{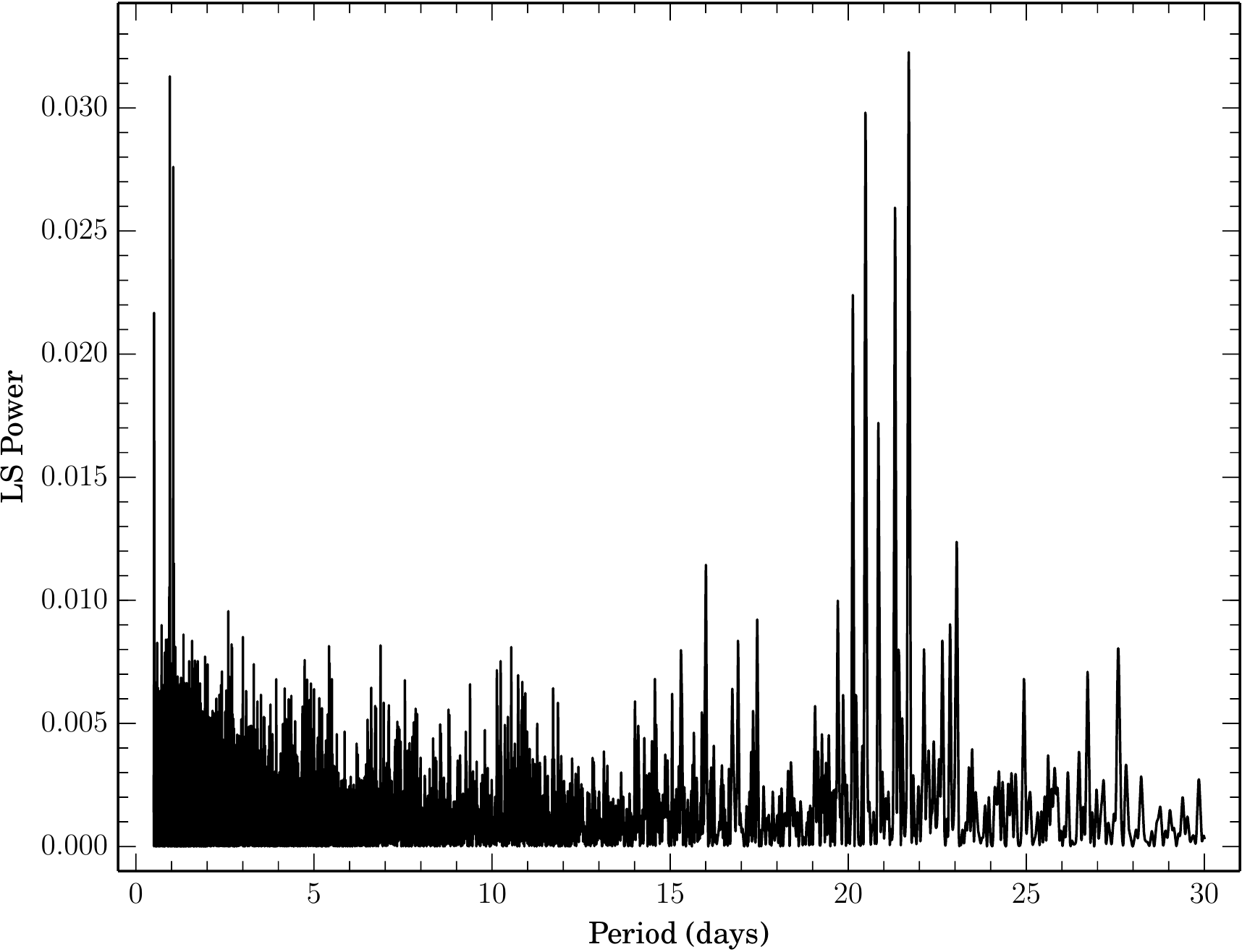}
    \caption{Lomb-Scargle periodogram between 0.5 and 30 days inferred from the smoothed OGLE I-band light curve shown in Fig.~\ref{OGLE_LC}.}
    \label{OGLE_LS}
\end{figure}

We followed a similar approach by smoothing the data, subtracted the smoothed curve in order to remove the long-term trends, and created a periodogram using the Lomb-Scargle algorithm (Fig.~\ref{OGLE_LS}). 
The periodogram between 0.5 and 30 days shows a series of peaks near 20--21 days and around 1 day. The six highest peaks of similar strength are at 0.953, 1.049, 20.133, 20.481, 21.314 and 21.705 days. The periods of all these peaks can be related to each other as aliases with periods of $\sim$364 days or $\sim$1180 days, the latter being caused by the outburst period. Short periods around one day are believed to be caused by non-radial pulsations (NRPs), a phenomenon commonly observed in Be stars \citep[see e.g.][]{2013A&ARv..21...69R}. Therefore, we interpret the 21.68 day period reported by \citet{Schurch2011} (consistent with our peak at 21.705 days) as likely being an alias of an NRP period close to one day. 

\citet{Maggi2014} suggested an orbital period of 1180 days for Swift\,J010745.0$-$722740, based on two strong outbursts seen in the OGLE light curve of this Be/X-ray binary in the SMC. No further optical outburst could be seen so far, but an X-ray detection in April 2017 at the time expected for the next outburst \citep{2017ATel10253....1V} confirmed the outburst period. 3XMM\,J051259.8$-$682640 in the Large Magellanic Cloud showed three remarkable dips in its 15 year OGLE light curve, suggesting a possible 1350 day orbital period \citep{Haberl2017}. The seven regular outbursts observed every $\sim$1163 days and the aliasing effects of a shorter period and $\sim$1180 days seen from \CXOJ\ might indicate the orbital period ($\sim$1160--1180 days) of the system after all and it could be the third Be-XRB with an orbital period longer than 1000 days. The outbursts would be caused by the perigee passage of the neutron star and are not expected to be strictly periodic due to long-term variations of the circum-stellar disc around the Be star.
Finally, we note that the largest orbital period in an Be-XRB systems is measured in PSR\,J2032+4127/MT91\,213, where the orbital period is $\sim$50 yr \citep{2015MNRAS.451..581L}, and its last periastron passage was in 2017 \citep{2018MNRAS.474L..22P,2017MNRAS.464.1211H}. Although, no major outburst was observed during the 2017 periastron passage of PSR\,J2032+4127, the discovery of the system demonstrates the existence of more Be-XRBs with very high orbital periods.

\section{Conclusions}\label{conclusions}
In this paper we present 1 Ms of new \NuSTAR\ observations of three fields in the Small Magellanic Cloud, including a catalog with 10 sources with greater than a 3$\sigma$ significance and 40 sources with upper limits on the source count rate.

We detected point sources down to a 3$\sigma$ luminosity limit of $10^{35}$ erg s$^{-1}$ in the 4--25 keV band, the lowest point source luminosity limit of any nearby galaxy observed with \NuSTAR. This detection sensitivity allowed us to analyze lower luminosity XRBs in the Small Magellanic Cloud than has been possible in other nearby galaxies. 

We used X-ray colors and luminosities to classify XRBs by compact-object type, black hole or neutron star, and to further subdivide black holes by accretion state and neutron stars as pulsars or low magnetic field neutron stars. We identified four sources as strongly variable both when we compare our \NuSTAR\ observations to archival \XMM\ observations as well as between epochs of our observations. 

We confirmed pulse periods for the seven known pulsars in our 3$\sigma$ sample (1728, 1677, 1701, 1705, 1666, 2012, 2035) using epoch folding for each source during each individual observation. We did not observe pulsations for the two HMXBs in our 3$\sigma$ sample that do not have confirmed pulse periods in the literature (1731, 1745).

We also present the first observations of periodic pulsations from SXP305 (\CXOJ), a Be-XRB. We measured an X-ray pulse period of $305.69\pm0.16$ seconds. \CXOJ\ is located 0.5\arcsec\ from the measured position of source 1705 and was observed in outburst during the first observation of Field 2 (50311002002). We did not detect pulsations during the second two observations of Field 2 (50311002004, 50311002006) because the nearby pulsar SXP15.3 was in outburst and dominated the flux detected at the location of 1705. The likely orbital period for this system is $\sim$1160-1180 days, which we measured using optical light curves from OGLE. 

We note that several low luminosity sources that are associated with confirmed pulsars fall in regions of the diagnostic diagrams consistent with multiple compact-object types. We raise questions about the apparent spectral differences of SMC pulsars as compared to the Milky Way pulsars that were used to create the diagnostics. Further work on \NuSTAR\ spectroscopic analysis for the sources in this catalog will be presented in V. Antoniou, et al., in preparation. More detailed pulse timing analysis for bright accreting pulsars will be presented in future work.

\acknowledgements This research has made use of the \NuSTAR\ Data Analysis Software (NuSTARDAS) jointly developed by the ASI Science Data Center (ASDC, Italy) and CalTech. This work made use of data supplied by the UK Swift Science Data Centre at the University of Leicester. This research has made use of NASA's Astrophysics Data System Bibliographic Services. This research has made use of the SIMBAD database,
operated at CDS, Strasbourg, France. This research made use of Astropy, a community-developed core Python package for Astronomy \citep{astropy}. This research made use of the X-ray spectral timing tool, \verb!stingray!  and \verb!HENDRICS! \citep{Huppenkothen2019}. The OGLE project has received funding from the National Science Centre, Poland, grant MAESTRO 2014/14/A/ST9/00121 to AU.

\facilities{NuSTAR, Swift (XRT), Chandra, XMM-Newton}
\software{HEAsoft \citep[v6.23,v6.24][]{2014ascl.soft08004N}, FTOOLS \citep{HEASOFT}, NuSTARDAS, astropy \citep{astropy2013,astropy2018}, stingray and HENDRICS \citep{Huppenkothen2019}, XSPEC \citep[v12.10.0c][]{Arnaud1996} }


\setlength{\tabcolsep}{2.5pt}
\begin{longrotatetable}

\begin{deluxetable*}{lllllllllcccc}

\tabletypesize{\scriptsize}
\tablecaption{\NuSTAR\ SMC Source Catalog\label{nustar_table}}
\tablehead{
\colhead{Source$^{a}$} & 
\colhead{R.A.} & 
\colhead{Dec.} & 
\colhead{F Count Rate} & 
\colhead{S Count Rate} & 
\colhead{M Count Rate} & 
\colhead{H Count Rate} & 
\colhead{$\frac{(M-S)}{(M+S)}$} &
\colhead{$\frac{(H-M)}{(H+M)}$} & 
\colhead{Exp. Time} & 
\colhead{Bgd. Rate} & 
\colhead{Field} &
\colhead{Haberl} \\
\colhead{ID} & 
\colhead{} & 
\colhead{} & 
\colhead{[$\times10^{-3}\ s^{-1}$]} & 
\colhead{[$\times10^{-3}\ s^{-1}$]} & 
\colhead{[$\times10^{-3}\ s^{-1}$]} & 
\colhead{[$\times10^{-3}\ s^{-1}$]} & 
\colhead{} &
\colhead{} & 
\colhead{[ks]} & 
\colhead{[s$^{-1}$]} & 
\colhead{} &
\colhead{ID}\\ 
}
\startdata
1701 & 12.686279 & --73.268131 & $19.57^{0.49}_{0.48}$ & $6.34^{0.30}_{0.29}$ & $9.62^{0.32}_{0.31}$ & $3.18^{0.20}_{0.19}$ & $0.20^{0.03}_{0.03}$ & $-0.48^{0.03}_{0.03}$ & 309 & 1.20 & 2 & 47 \\
1745 & 12.738000 & --73.168813 & $41.60^{0.71}_{0.70}$ & $14.21^{0.43}_{0.41}$ & $20.60^{0.46}_{0.45}$ & $6.89^{0.33}_{0.32}$ & $0.17^{0.02}_{0.02}$ & $-0.49^{0.02}_{0.02}$ & 207 & 1.40 & 2 & 84 \\
2012 & 13.023879 & --72.434562 & $101.21^{0.89}_{0.91}$ & $38.00^{0.60}_{0.59}$ & $51.60^{0.64}_{0.61}$ & $12.07^{0.34}_{0.33}$ & $0.15^{0.01}_{0.01}$ & $-0.59^{0.01}_{0.01}$ & 259 & 1.20 & 3 & 9 \\
1705$^{b}$ & 13.064458 & --73.320945 & $951.58^{3.70}_{3.60}$ & $244.96^{2.00}_{1.90}$ & $482.76^{2.60}_{2.50}$ & $221.40^{1.90}_{1.90}$ & $0.32^{0.00}_{0.00}$ & $-0.36^{0.00}_{0.00}$ & 107 & 1.20 & 2 & 93/16 \\
1666 & 13.349638 & --72.454316 & $27.47^{0.58}_{0.56}$ & $8.95^{0.32}_{0.30}$ & $13.63^{0.37}_{0.36}$ & $4.39^{0.26}_{0.26}$ & $0.19^{0.02}_{0.02}$ & $-0.49^{0.02}_{0.02}$ & 192 & 1.20 & 3 & 30 \\
2035 & 13.480736 & --72.445979 & $20.82^{0.71}_{0.68}$ & $7.00^{0.40}_{0.38}$ & $9.44^{0.44}_{0.42}$ & $4.24^{0.39}_{0.36}$ & $0.16^{0.03}_{0.03}$ & $-0.34^{0.04}_{0.04}$ & 151 & 1.20 & 3 & 22 \\
2052 & 13.713261 & --72.442148 & $22.79^{2.20}_{2.20}$ & $10.24^{1.30}_{1.20}$ & $14.92^{1.50}_{1.50}$ & $0.00^{0.84}_{--}$ & $0.17^{0.08}_{0.08}$ & $-0.94^{0.10}_{--}$ & 260 & 1.10 & 1 & -- \\
1677 & 13.734407 & --72.446681 & $906.22^{3.10}_{3.00}$ & $242.76^{1.70}_{1.70}$ & $449.76^{2.10}_{2.00}$ & $190.81^{1.10}_{1.10}$ & $0.30^{0.00}_{0.00}$ & $-0.41^{0.00}_{0.00}$ & 285 & 1.10 & 1 & 23 \\
1728 & 13.896565 & --72.485134 & $26.68^{0.63}_{0.61}$ & $9.02^{0.39}_{0.29}$ & $13.12^{0.42}_{0.40}$ & $4.30^{0.26}_{0.25}$ & $0.19^{0.03}_{0.03}$ & $-0.49^{0.03}_{0.03}$ & 185 & 1.10 & 1 & 55 \\
1731 & 14.078402 & --72.467787 & $8.56^{0.39}_{0.37}$ & $2.60^{0.22}_{0.21}$ & $4.14^{0.24}_{0.23}$ & $1.76^{0.20}_{0.19}$ & $0.20^{0.04}_{0.04}$ & $-0.37^{0.05}_{0.05}$ & 176 & 1.10 & 1 & 117 \\
\hline \\
1695 & 12.294759 & --73.288091 & <2.0 & <0.7 & <0.9 & <0.6 & $0.06^{0.17}_{0.18}$ & $-0.47^{0.25}_{--}$ & 144 & 1.20 & 2 & -- \\
2111 & 12.338442 & --73.294647 & <0.6 & <0.4 & <0.2 & <0.2 & $-0.38^{0.37}_{--}$ & $-1.00^{1.10}_{--}$ & 220 & 1.20 & 2 & -- \\
2115 & 12.488716 & --73.282034 & <0.0 & <0.2 & <0.0 & <0.0 & $--^{--}_{--}$ & $--^{--}_{--}$ & 307 & 1.20 & 2 & -- \\
2118 & 12.522018 & --73.196036 & <0.5 & <0.1 & <0.2 & <0.2 & $0.22^{0.23}_{0.22}$ & $-0.30^{0.24}_{--}$ & 247 & 1.70 & 2 & -- \\
2119 & 12.527318 & --73.263826 & <0.0 & <0.1 & <0.0 & <0.0 & $-0.02^{--}_{--}$ & $-1.00^{1.60}_{--}$ & 323 & 1.20 & 2 & -- \\
2121 & 12.598875 & --73.305656 & <0.7 & <0.4 & <0.4 & <0.1 & $0.13^{0.12}_{0.12}$ & $-0.44^{0.15}_{0.17}$ & 330 & 1.20 & 2 & -- \\
2123 & 12.675824 & --73.360578 & <0.1 & <0.3 & <0.0 & <0.1 & $--^{--}_{--}$ & $--^{--}_{--}$ & 272 & 1.10 & 2 & -- \\
1702 & 12.687766 & --73.261037 & <4.5 & <2.0 & <2.2 & <0.6 & $0.06^{0.09}_{0.09}$ & $-0.51^{0.10}_{0.10}$ & 322 & 1.20 & 2 & -- \\
2124 & 12.687912 & --73.255423 & <0.3 & <0.1 & <0.3 & <0.2 & $--^{--}_{--}$ & $--^{--}_{--}$ & 302 & 1.20 & 2 & -- \\
2125 & 12.699929 & --73.304988 & <1.2 & <0.7 & <0.5 & <0.2 & $-0.04^{0.11}_{0.11}$ & $-0.29^{0.14}_{0.15}$ & 343 & 1.20 & 2 & 45 \\
2127 & 12.731243 & --73.343496 & <1.4 & <0.5 & <0.8 & <0.3 & $0.15^{0.17}_{0.16}$ & $-0.38^{0.19}_{0.24}$ & 298 & 1.10 & 2 & -- \\
2129 & 12.746321 & --73.348747 & <1.4 & <0.3 & <0.4 & <0.8 & $-0.07^{0.37}_{--}$ & $0.44^{0.23}_{0.23}$ & 288 & 1.10 & 2 & -- \\
1864 & 12.762729 & --73.357891 & <0.6 & <0.1 & <0.2 & <0.6 & $0.53^{--}_{--}$ & $0.67^{--}_{--}$ & 274 & 1.10 & 2 & -- \\
1865 & 12.763887 & --73.360729 & <0.5 & <0.0 & <0.2 & <0.5 & $--^{--}_{--}$ & $0.31^{1.30}_{--}$ & 269 & 1.10 & 2 & -- \\
2134 & 12.915870 & --73.300610 & <0.7 & <0.4 & <0.4 & <0.2 & $0.52^{0.09}_{0.13}$ & $0.27^{0.03}_{0.09}$ & 298 & 1.20 & 2 & -- \\
1662 & 12.972087 & --72.530245 & <0.3 & <0.2 & <0.3 & <0.1 & $0.18^{0.30}_{0.29}$ & $-0.15^{0.29}_{0.32}$ & 275 & 1.20 & 3 & 11 \\
2137 & 13.008025 & --73.234996 & <0.1 & <0.2 & <0.2 & <0.0 & $0.20^{0.15}_{0.15}$ & $0.00^{0.15}_{0.14}$ & 239 & 1.00 & 2 & -- \\
94 & 13.038505 & --72.431364 & <0.2 & <0.2 & <0.2 & <0.2 & $0.95^{--}_{--}$ & $-0.99^{--}_{--}$ & 260 & 1.20 & 3 & -- \\
96 & 13.066980 & --72.433882 & <0.2 & <0.2 & <0.1 & <0.1 & $-0.97^{--}_{--}$ & $0.97^{--}_{--}$ & 271 & 1.20 & 3 & -- \\
1663 & 13.140267 & --72.410466 & <0.8 & <0.6 & <0.6 & <0.1 & $0.13^{0.25}_{0.23}$ & $-0.52^{0.29}_{--}$ & 244 & 1.20 & 3 & -- \\
1664 & 13.146238 & --72.421389 & <0.5 & <0.3 & <0.3 & <0.1 & $-0.07^{0.67}_{--}$ & $0.16^{0.43}_{0.42}$ & 265 & 1.20 & 3 & 94 \\
2017 & 13.243471 & --72.435198 & <0.1 & <0.1 & <0.2 & <0.0 & $--^{--}_{--}$ & $--^{--}_{--}$ & 278 & 1.20 & 3 & -- \\
2024 & 13.382805 & --72.446137 & <0.7 & <0.2 & <0.3 & <0.6 & $0.61^{--}_{--}$ & $0.56^{--}_{0.34}$ & 178 & 1.20 & 3 & -- \\
1668 & 13.407895 & --72.402495 & <0.6 & <0.2 & <0.7 & <0.1 & $0.51^{--}_{0.28}$ & $-0.33^{0.26}_{--}$ & 137 & 1.20 & 3 & -- \\
1670 & 13.468735 & --72.533149 & <2.3 & <0.7 & <0.9 & <0.9 & $0.08^{0.13}_{0.13}$ & $0.10^{0.12}_{0.13}$ & 166 & 1.20 & 3 & -- \\
1403 & 13.478602 & --72.456238 & <3.9 & <0.8 & <1.6 & <1.9 & $0.27^{0.17}_{0.16}$ & $0.15^{0.12}_{0.11}$ & 155 & 1.20 & 3 & -- \\
2047 & 13.619948 & --72.511644 & <0.2 & <0.3 & <0.1 & <0.1 & $-0.99^{--}_{--}$ & $-1.00^{--}_{--}$ & 227 & 1.10 & 1 & -- \\
1722 & 13.621017 & --72.518590 & <0.2 & <0.2 & <0.1 & <0.2 & $-0.04^{0.33}_{0.33}$ & $-0.08^{0.43}_{--}$ & 219 & 1.10 & 1 & -- \\
1673 & 13.654902 & --72.443810 & <0.1 & <0.1 & <0.2 & <0.0 & $--^{--}_{--}$ & $--^{--}_{--}$ & 264 & 1.10 & 1 & -- \\
1674 & 13.688963 & --72.399620 & <1.0 & <0.4 & <0.7 & <0.2 & $--^{--}_{--}$ & $-0.33^{0.58}_{--}$ & 259 & 1.10 & 1 & -- \\
1675 & 13.693395 & --72.423048 & <0.7 & <1.1 & <0.3 & <0.2 & $-0.99^{--}_{--}$ & $-0.97^{--}_{--}$ & 279 & 1.10 & 1 & 63 \\
1676 & 13.704129 & --72.429160 & <1.8 & <0.7 & <0.9 & <0.6 & $0.11^{--}_{--}$ & $-0.20^{--}_{--}$ & 273 & 1.10 & 1 & -- \\
1678 & 13.738063 & --72.506046 & <1.2 & <0.3 & <0.8 & <0.4 & $0.86^{--}_{0.28}$ & $-0.36^{0.21}_{0.24}$ & 268 & 1.10 & 1 & -- \\
1679 & 13.768155 & --72.375208 & <0.2 & <0.3 & <0.4 & <0.0 & $0.21^{0.76}_{0.40}$ & $-1.00^{0.26}_{--}$ & 247 & 1.10 & 1 & -- \\
1680 & 13.782369 & --72.378083 & <0.1 & <0.3 & <0.1 & <0.0 & $-0.09^{0.78}_{--}$ & $-1.00^{0.81}_{--}$ & 253 & 1.10 & 1 & 111 \\
1724 & 13.826634 & --72.523471 & <2.3 & <0.7 & <0.8 & <1.0 & $0.14^{0.13}_{0.13}$ & $0.05^{0.12}_{0.12}$ & 172 & 1.10 & 1 & -- \\
1726 & 13.894533 & --72.476154 & <3.0 & <1.3 & <1.8 & <0.3 & $0.33^{0.21}_{0.18}$ & $-0.72^{0.18}_{--}$ & 198 & 1.10 & 1 & -- \\
1760 & 13.923780 & --72.448851 & <0.1 & <0.1 & <0.1 & <0.1 & $1.00^{--}_{--}$ & $0.17^{--}_{2.90}$ & 203 & 1.10 & 1 & -- \\
2232 & 13.985326 & --72.492418 & <0.6 & <0.3 & <0.2 & <0.3 & $-0.19^{--}_{--}$ & $0.34^{0.59}_{0.46}$ & 187 & 1.10 & 1 & -- \\
2237 & 14.094049 & --72.507746 & <0.6 & <0.2 & <0.2 & <0.5 & $-0.64^{0.99}_{--}$ & $0.79^{0.65}_{--}$ & 156 & 1.10 & 1 & -- \\
\enddata
\tablecomments{\NuSTAR\ source catalog with count rates and hardness ratios from combining all observations of each field. Bands used are as follows: $S=4-6$ keV, $M=6-12$ keV, $H=12-25$ keV, $F=4-25$ keV. Sources listed above the horizontal line were detected at a 3$\sigma$ significance level, all sources listed after the horizontal line are upper limits. Entries listed as ``--'' indicate missing values. Haberl ID indicates listed identification number in \citet{Haberl2016}. Background count rate was fit using the methods described in Section \ref{bgd_fitting}. Exposure times listed include both FPMA and FPMB telescope images, with the exception of the two FPMB telescope images that were omitted from source extraction due to stray light contamination in Fields 1 and 3. \newline
$^{a}$ We use source IDs are from \citet{Antoniou2019} as the source positions from that catalog were used as the priors on source positions in our PSF fitting routine to extract count rates and hardness ratios.
\newline$^{b}$ \NuSTAR\ source 1705 is a combination of two sources, SXP15.3 and the newly confirmed pulsar SXP305. See \S \ref{new_pulsar} for more details.}
\end{deluxetable*}
\end{longrotatetable}

\begin{longrotatetable}
\begin{deluxetable*}{lllllllllcc}
\tabletypesize{\scriptsize}
\tablecaption{Count Rate and Hardness Ratios By Observation\label{nustar_time}}
\tablehead{
\colhead{Source$^{a}$} & 
\colhead{R.A.} & 
\colhead{Dec.} & 
\colhead{F Count Rate} & 
\colhead{S Count Rate} & 
\colhead{M Count Rate} & 
\colhead{H Count Rate} & 
\colhead{$\frac{(M-S)}{(M+S)}$} &
\colhead{$\frac{(H-M)}{(H+M)}$} & 
\colhead{Obs.} & 
\colhead{Date} \\
\colhead{ID} & 
\colhead{} & 
\colhead{} & 
\colhead{[$\times10^{-3}\ s^{-1}$]} & 
\colhead{[$\times10^{-3}\ s^{-1}$]} & 
\colhead{[$\times10^{-3}\ s^{-1}$]} & 
\colhead{[$\times10^{-3}\ s^{-1}$]} & 
\colhead{} &
\colhead{} &
\colhead{ID} &
\colhead{yyyy-mm-dd} \\ 
}
\startdata
1701 & 12.686279 & --73.268131 & $18.65^{0.61}_{0.60}$ & $6.08^{0.37}_{0.36}$ & $8.95^{0.39}_{0.39}$ & $3.22^{0.26}_{0.25}$ & $0.18^{0.04}_{0.04}$ & $-0.45^{0.03}_{0.03}$ & 50311002002 & 2017-03-12 \\
1701 & 12.686279 & --73.268131 & $22.83^{0.91}_{0.91}$ & $7.85^{0.62}_{0.57}$ & $11.74^{0.50}_{0.49}$ & $3.03^{0.32}_{0.29}$ & $0.20^{0.04}_{0.04}$ & $-0.54^{0.04}_{0.04}$ & 50311002004 & 2017-07-19 \\
1701 & 12.686279 & --73.268131 & $30.15^{1.40}_{1.30}$ & $9.49^{0.86}_{0.78}$ & $14.44^{0.90}_{0.84}$ & $5.11^{0.75}_{0.68}$ & $0.19^{0.05}_{0.05}$ & $-0.44^{0.05}_{0.04}$ & 50311002006 & 2017-08-09 \\
1745 & 12.738000 & --73.168813 & $35.10^{0.81}_{0.79}$ & $12.24^{0.49}_{0.47}$ & $17.57^{0.53}_{0.52}$ & $5.47^{0.38}_{0.36}$ & $0.17^{0.02}_{0.02}$ & $-0.51^{0.02}_{0.02}$ & 50311002002 & 2017-03-12 \\
1745 & 12.738000 & --73.168813 & $65.72^{1.80}_{1.70}$ & $21.71^{1.10}_{1.00}$ & $32.30^{1.20}_{1.10}$ & $11.99^{0.86}_{0.81}$ & $0.18^{0.03}_{0.03}$ & $-0.46^{0.03}_{0.03}$ & 50311002004 & 2017-07-19 \\
1745 & 12.738000 & --73.168813 & $32.84^{3.70}_{3.30}$ & $11.72^{2.20}_{1.90}$ & $15.57^{2.50}_{2.20}$ & $5.98^{1.80}_{1.40}$ & $0.16^{0.09}_{0.09}$ & $-0.58^{0.11}_{0.10}$ & 50311002006 & 2017-08-09 \\
2012 & 13.023879 & --72.434562 & $118.05^{1.50}_{1.20}$ & $42.47^{0.95}_{0.94}$ & $59.76^{1.10}_{1.00}$ & $15.88^{0.64}_{0.45}$ & $0.17^{0.01}_{0.01}$ & $-0.58^{0.01}_{0.01}$ & 50311003002 & 2017-05-03 \\
2012 & 13.023879 & --72.434562 & $94.74^{1.30}_{1.30}$ & $35.54^{0.85}_{0.79}$ & $48.33^{0.88}_{0.86}$ & $11.29^{0.51}_{0.50}$ & $0.14^{0.01}_{0.01}$ & $-0.58^{0.01}_{0.01}$ & 50311003004 & 2017-08-07 \\
1705$^{a}$ & 13.064458 & --73.320945 & $60.71^{1.60}_{1.50}$ & $18.89^{0.90}_{0.86}$ & $32.01^{1.10}_{1.00}$ & $9.54^{0.76}_{0.71}$ & $0.26^{0.03}_{0.03}$ & $-0.52^{0.03}_{0.03}$ & 50311002002 & 2017-03-12 \\
1705$^{a}$ & 13.064458 & --73.320945 & $1188.09^{5.30}_{5.00}$ & $313.94^{2.80}_{2.70}$ & $595.60^{3.50}_{3.50}$ & $268.75^{2.60}_{2.60}$ & $0.31^{--}_{--}$ & $-0.35^{--}_{--}$ & 50311002004 & 2017-07-19 \\
1705$^{a}$ & 13.064458 & --73.320945 & $1749.18^{20.00}_{15.00}$ & $466.64^{14.00}_{6.60}$ & $889.38^{15.00}_{9.50}$ & $383.44^{8.60}_{8.40}$ & $0.31^{0.01}_{0.01}$ & $-0.39^{0.01}_{0.01}$ & 50311002006 & 2017-08-09 \\
1666 & 13.349638 & --72.454316 & $25.97^{0.72}_{0.73}$ & $8.13^{0.42}_{0.40}$ & $13.06^{0.48}_{0.46}$ & $4.11^{0.33}_{0.32}$ & $0.20^{0.03}_{0.03}$ & $-0.54^{0.03}_{0.03}$ & 50311003002 & 2017-05-03 \\
1666 & 13.349638 & --72.454316 & $22.15^{0.64}_{0.65}$ & $7.36^{0.43}_{0.42}$ & $10.71^{0.43}_{0.11}$ & $4.26^{0.38}_{0.37}$ & $0.18^{0.03}_{0.03}$ & $-0.40^{0.03}_{0.03}$ & 50311003004 & 2017-08-07 \\
2035 & 13.480736 & --72.445979 & $6.32^{0.66}_{0.59}$ & $1.50^{0.34}_{0.31}$ & $2.50^{0.40}_{0.35}$ & $2.27^{0.40}_{0.36}$ & $0.21^{0.07}_{0.07}$ & $-0.18^{0.08}_{0.08}$ & 50311003002 & 2017-05-03 \\
2035 & 13.480736 & --72.445979 & $33.97^{1.60}_{1.50}$ & $11.60^{1.00}_{0.85}$ & $15.75^{0.99}_{1.20}$ & $8.41^{1.00}_{0.85}$ & $0.16^{0.03}_{0.03}$ & $-0.29^{0.03}_{0.03}$ & 50311003004 & 2017-08-07 \\
2052 & 13.713261 & --72.442148 & $<0.32$ & $<0.61$ & $<0.44$ & $<0.17$ & $0.97^{--}_{--}$ & $0.38^{--}_{--}$ & 50311001002 & 2017-04-24 \\
2052 & 13.713261 & --72.442148 & $<0.15$ & $<0.25$ & $<0.12$ & $<0.10$ & $0.95^{--}_{--}$ & $1.00^{--}_{--}$ & 50311001004 & 2017-08-12 \\
1677 & 13.734407 & --72.446681 & $1867.01^{5.00}_{5.00}$ & $495.88^{2.80}_{2.70}$ & $929.93^{3.40}_{3.40}$ & $403.31^{2.30}_{2.30}$ & $0.30^{--}_{--}$ & $-0.40^{--}_{--}$ & 50311001002 & 2017-04-24 \\
1677 & 13.734407 & --72.446681 & $152.74^{1.50}_{1.40}$ & $51.39^{0.91}_{0.87}$ & $76.33^{0.99}_{0.96}$ & $23.57^{0.58}_{0.56}$ & $0.19^{0.01}_{0.01}$ & $-0.52^{0.01}_{0.01}$ & 50311001004 & 2017-08-12 \\
1728 & 13.896565 & --72.485134 & $31.89^{0.96}_{0.90}$ & $10.68^{0.60}_{0.57}$ & $16.14^{0.63}_{0.61}$ & $4.74^{0.35}_{0.34}$ & $0.22^{0.03}_{0.03}$ & $-0.53^{0.03}_{0.03}$ & 50311001002 & 2017-04-24 \\
1728 & 13.896565 & --72.485134 & $20.94^{0.65}_{0.63}$ & $7.43^{0.41}_{0.39}$ & $9.86^{0.47}_{0.46}$ & $3.22^{0.28}_{0.26}$ & $0.15^{0.03}_{0.03}$ & $-0.51^{0.03}_{0.03}$ & 50311001004 & 2017-08-12 \\
1731 & 14.078402 & --72.467787 & $13.19^{0.56}_{0.54}$ & $3.94^{0.32}_{0.30}$ & $6.12^{0.35}_{0.34}$ & $2.86^{0.28}_{0.26}$ & $0.22^{0.04}_{0.04}$ & $-0.38^{0.04}_{0.04}$ & 50311001002 & 2017-04-24 \\
1731 & 14.078402 & --72.467787 & $2.08^{0.42}_{0.39}$ & $0.57^{0.25}_{0.21}$ & $1.33^{0.27}_{0.23}$ & $0.22^{0.21}_{--}$ & $0.16^{0.15}_{0.14}$ & $-0.45^{0.17}_{0.20}$ & 50311001004 & 2017-08-12 \\
\enddata
\tablecomments{Catalog of count rates and hardness ratio for 3$\sigma$ sources broken down by observation. Bands used are as follows: $S=4-6$ keV, $M=6-12$ keV, $H=12-25$ keV, $F=4-25$ keV. We note that source 2052 was not detected when each observation was reduced separately, but the count rates and hardness ratios from the combined observations can be found in Table \ref{nustar_table}.\newline
$^{a}$ We use source IDs are from \citet{Antoniou2019} as the source positions from that catalog were used as the priors on source positions in our PSF fitting routine to extract count rates and hardness ratios.
\newline$^{a}$ \NuSTAR\ source 1705 is a combination of two sources, SXP15.3 and the newly confirmed pulsar SXP305. The new pulsar, SXP305, was in outburst during the first observation and the previously known pulsar SXP15.3 was in outburst during the second two observations. See \S \ref{new_pulsar} for more details.}
\end{deluxetable*}
\end{longrotatetable}
\clearpage

\begin{deluxetable*}{cccccc}
\tablecaption{\swift\ Observations List \label{obs_list2}}
\tablehead{
\colhead{Obs. ID} &  
\colhead{R.A.} & 
\colhead{Dec.} & 
\colhead{Field} & 
\colhead{Exp.} & 
\colhead{Date (start)} \\ 
\colhead{} &  
\colhead{(J2000)} & 
\colhead{(J2000)} & 
\colhead{ID} & 
\colhead{Time [s]} & 
\colhead{MJD}
}
\startdata
00088082001 &  13.87418 &  -72.4405 &   1 &  2419.87 &  57867.79688  \\
00088082002 &  13.87418 &  -72.4405 &   1 &  12089.4 &  57871.25364  \\
00088082003 &  13.93822 &  -72.4248 &   1 &  9322.35 &  57976.09448  \\
00088082004 &  13.92335 &  -72.4361 &   1 &  1882.95 &  57977.53664  \\
00088083001 &  12.69375 &  -73.2741 &   2 &  7202.19 &  57824.08328  \\
00088083002 &  12.69375 &  -73.2741 &   2 &  7272.09 &  57826.00306  \\
00088083003 &  12.69375 &  -73.2741 &   2 &  7272.09 &  57827.06391  \\
00088083004 &  12.89389 &  -73.2506 &   2 &  983.931 &  57959.94944  \\
00088083005 &  12.74704 &  -73.2501 &   2 &  6692.72 &  57960.28014  \\
00088083006 &  12.71321 &  -73.2655 &   2 &  9324.84 &  57975.61163  \\
00088032001 &  13.21749 &  -72.4804 &   3 &  5034.55 &  57876.30592  \\
00088032002 &  13.21749 &  -72.4804 &   3 &  5651.39 &  57877.11145  \\
00088032003 &  13.21749 &  -72.4804 &   3 &  4247.88 &  57878.02015  \\
00088032004 &  13.25689 &  -72.5004 &   3 &  10076.5 &  57972.10376  \\
00088032005 &  13.24207 &  -72.4562 &   3 &  5526.52 &  57973.02310  \\
\enddata
\tablecomments{Observation IDs, positions, corresponding \NuSTAR\ field, exposure time, and observation date for \swift\ observations.}
\end{deluxetable*}

\begin{deluxetable*}{cccc}
\tabletypesize{\scriptsize}
\tablecaption{\swift\ 0.2-10 keV Count Rates For Each \textit{Swift} Observation \label{swift_data}}
\tablehead{
\colhead{Source ID} & 
\colhead{Count Rate [s$^{-1}$]} & 
\colhead{Obs. ID} & 
\colhead{Date [yyyy-mm-dd]}
}
\startdata
1701 & $0.007\pm0.001$ & 88083001 & 2017-03-12 \\
1701 & $0.010\pm0.001$ & 88083002 & 2017-03-14 \\
1701 & $0.006\pm0.001$ & 88083003 & 2017-03-15 \\
1701 & $0.011\pm0.004$ & 88083004 & 2017-07-25 \\
1701 & $0.011\pm0.002$ & 88083005 & 2017-07-26 \\
1701 & $0.010\pm0.001$ & 88083006 & 2017-08-10 \\
1745 & $0.017\pm0.002$ & 88083003 & 2017-03-15 \\
1745 & $0.017\pm0.002$ & 88083005 & 2017-07-26 \\
2012 & $0.054\pm0.004$ & 88032001 & 2017-05-03 \\
2012 & $0.045\pm0.003$ & 88032002 & 2017-05-04 \\
2012 & $0.037\pm0.003$ & 88032003 & 2017-05-05 \\
2012 & $0.032\pm0.002$ & 88032004 & 2017-08-07 \\
2012 & $0.028\pm0.003$ & 88032005 & 2017-08-08 \\
1705$^{a}$ & $0.702\pm0.028$ & 88083004 & 2017-07-25 \\
1666 & $0.013\pm0.002$ & 88032001 & 2017-05-03 \\
1666 & $0.014\pm0.002$ & 88032002 & 2017-05-04 \\
1666 & $0.009\pm0.002$ & 88032003 & 2017-05-05 \\
1666 & $0.010\pm0.001$ & 88032004 & 2017-08-07 \\
1666 & $0.005\pm0.001$ & 88032005 & 2017-08-08 \\
2035 & $0.015\pm0.001$ & 88032004 & 2017-08-07 \\
2035 & $0.017\pm0.002$ & 88032005 & 2017-08-08 \\
1673 & $0.002\pm0.001$ & 88082003 & 2017-08-11 \\
1677 & $0.431\pm0.014$ & 88082001 & 2017-04-24 \\
1677 & $0.043\pm0.002$ & 88082003 & 2017-08-11 \\
1677 & $0.038\pm0.005$ & 88082004 & 2017-08-12 \\
1728 & $0.010\pm0.002$ & 88082001 & 2017-04-24 \\
1728 & $0.011\pm0.001$ & 88082002 & 2017-04-28 \\
1728 & $0.008\pm0.001$ & 88082003 & 2017-08-11 \\
1728 & $0.011\pm0.003$ & 88082004 & 2017-08-12 \\
1731 & $0.004\pm0.001$ & 88082002 & 2017-04-28 \\
\enddata
\tablecomments{Source count rates for quasi-simultaneous \swift\ observations that were used to create Figure \ref{swift_Comparison}. \swift\ count rates are for the full 0.2-10 keV energy band. Sources are listed by their source IDs in Table \ref{nustar_table}. For more information on \swift\ count rate fitting, see \S \ref{swift_observations}. For exposure times for each \swift\ observation, see Table \ref{obs_list2}.
\newline$^{a}$ \NuSTAR\ source 1705 is a combination of two sources, SXP15.3 and the newly confirmed pulsar SXP305. See \S \ref{new_pulsar} for more details.}
\end{deluxetable*}

\begin{deluxetable*}{cccc}
\tabletypesize{\normalsize}
\tablecaption{Pulsation Periods \label{pulse_periods}}
\tablehead{\colhead{Source ID} &
\colhead{Measured Pulse Period [s]} & 
\colhead{Pulsed Fraction} & 
\colhead{Observation Start [MJD]}
}
\startdata
1666A (SXP138) & 140.73$\pm$0.04 & 0.44$\pm$0.04 &  57876.12 \\
1666B (SXP138) & 140.85$\pm$0.05 & 0.49$\pm$0.06 &  57972.12 \\
\hline
1677A (SXP59.0) & 58.863$\pm$0.007 &  0.562$\pm$0.005 &  57867.07\\
1677B (SXP59.0) & 58.799$\pm$0.010 & 0.38$\pm$0.03 &  57977.09\\
\hline
1701A (SXP323) & 316.19$\pm$0.16 & 0.5$\pm$0.05 &  57824.07\\
1701B (SXP323) & 316.26$\pm$0.24 & 0.47$\pm$0.06 &  57953.59\\
1701C (SXP323) & 316.1$\pm$0.9 & 0.58$\pm$0.09 &  57974.96\\
\hline
1705A (SXP305) & 305.69$\pm$0.15 (new pulsar) & 0.55$\pm$0.05  & 57824.07\\
1705B (SXP15.3) & 15.2822$\pm$0.0005 & 0.251$\pm$0.013 &  57953.60\\
1705C (SXP15.3) & 15.2738$\pm$0.0016  & 0.32$\pm$0.03 &  57974.96\\
\hline
1728A (SXP645) & 647.2$\pm$0.8 & 0.24$\pm$0.04 & 57867.07\\
1728B (SXP645) & -- &  <0.32 &  57977.09\\
\hline
2012A (SXP7.77) & 7.76923$\pm$0.00011 & 0.45$\pm$0.02 &  57876.12\\
2012B (SXP7.77) & 7.76903$\pm$0.00014 & 0.47$\pm$0.04 &  57972.12\\
\hline
2035A (SXP46.6) & 58.826$\pm$0.006* & 0.31$\pm$0.06 &  57876.12\\
2035B (SXP46.6) & 45.981$\pm$0.005 &  0.32$\pm$0.06 &  57972.12\\
\enddata
\tablecomments{Measured pulse periods for pulsars in 3$\sigma$ source sample. Pulse periods were measured using epoch folding method, see \S \ref{timing} for an overview of methodology. The measured pulse period for each source during each observation are listed separately. 
Pulsed Fractions are given for detections, based on a folded pulse profile with 16 phase bins, upper limits are given for non detections. 
The suffix A, B, C correspond to the first, second, and third observations of the field containing that source. See Table \ref{obs_list} for an overview of observations.\newline
*Source 2035 is 4.6 arcmin away from 1677 (SXP\,59). During the first observation of Field 3, no point source (i.e. SXP\,46.6) was visible in NuSTAR image, we only see background contamination. The fit pulse period reflects background contamination from source SXP\,59.}
\end{deluxetable*}

\setlength{\tabcolsep}{5pt}
\begin{longrotatetable}
\begin{deluxetable*}{lllllcl}

\tabletypesize{\scriptsize}

\tablecaption{\NuSTAR\ SMC Source Classifications\label{class_table}}
\tablehead{
\colhead{NuSTAR} &  
\colhead{$L_{4-25keV}$} &
\colhead{Diagnostic Diagram} & 
\colhead{Haberl} & 
\colhead{Sep.} &
\colhead{Simbad} &
\colhead{Properties from}\\
\colhead{ID} &  
\colhead{[$\times 10^{35}\ erg\ s^{-1}$]} &
\colhead{Classification} & 
\colhead{ID} & 
\colhead{[$\prime \prime$]} & 
\colhead{Name} &
\colhead{\citet{Haberl2016}}\\
}
\startdata
1728 & 5.86 & PUL, HBH, IBH, NS & 55 & 1.14 & SXP645, XMMU J005535.2$-$722906 & confirmed pulsar \\
1677 & 199 & PUL, IBH, HBH, NS & 23 & 1.07 & SXP59.0, RX J0054.9$-$7226 & confirmed pulsar \\
2052 & 5.01 & HBH$^{a}$ & -- & 1.71 & XMMU J005451.2$-$722630 & likely AGN \citep{Sturm2013} \\\vspace{-0.15cm}
1731 & 1.88 & HBH, PUL, IBH & 117 & 1.10 & XMMU J005618.8$-$722802 & emission line star optical counterpart \\
 & & & & & & within positional errors \\
1701 & 4.30 & HBH, IBH, PUL & 47 & 0.54 & SXP323, RX J0050.8$-$7316 & confirmed pulsar \\
1745 & 9.14 & PUL, HBH, IBH, NS & 84 & 0.83 & RX J0050.9$-$7310 & hard Be-XRB spectrum\\
1705$^{b}$ & 209 & PUL, HBH & 93/16 & 0.50 & SXP15.3 and SXP305 & confirmed pulsar and newly confirmed pulsar  \\
1666 & 6.04 & HBH, PUL, IBH, NS & 30 & 0.61 & SXP138, CXOU J005323.8$-$722715 & confirmed pulsar \\
2012 & 22.2 & NS, PUL, HBH, IBH & 9 & 0.45 & SXP7.77, SMC X-3 & confirmed pulsar \\
2035 & 4.58 & HBH, PUL & 22 & 0.93 & SXP46.6, XTE J0053$-$724 & confirmed pulsar \\
\enddata
\tablecomments{Table including \NuSTAR\ source ID, 4--25 keV luminosities, our classification of each source using the diagnostic diagrams (figures \ref{var_1666}--\ref{var_2035}), and ID, Simbad names and source properties listed in the \citet{Haberl2016} catalog. The column labeled ``Sep.'' indicates the separation between our source position and the position listed in \citet{Haberl2016} in arcseconds. Source classifications were determined by inspecting the source positions on the diagnostic hardness-intensity and hardness ratio diagrams during each individual epoch of observation (Figs. \ref{var_1666}-\ref{var_2035}). We did not use the hardness ratios and count rates from the merged observations because most sources changed full band luminosity and/or hardness ratios between epochs. For sources that lie in overlapping regions between compact-object types on the diagnostic diagrams, we list all possible classifications. We note that SMC \NuSTAR\ source 2052 did not match to any source in the \citet{Haberl2016} catalog within 5\arcsec\ because it is a likely background AGN, not a HMXB. All sources with $L_{4-25keV}<1.0\times 10^{36}\ erg\ s^{-1}$ are considered low luminosity sources, when compared with XRB populations studied in other nearby galaxies.
\newline $^{a}$ Source 2052 is classified using the diagnostic diagram created by combining all epochs of observation shown in Figure \ref{colorcolor} because it did not have count rate and hardness ratio measurements with well defined errors in each individual observation.
\newline$^{b}$ Source 1705 is a combination of two sources, SXP15.3 and the newly confirmed pulsar SXP305. See \S \ref{new_pulsar} for more details.}
\end{deluxetable*}
\end{longrotatetable}

\begin{deluxetable*}{cc}
\tablecaption{Flux Conversion Factors For Figure \ref{xmm} \label{conversion}}
\tablehead{
\colhead{Model} &
\colhead{Conversion Factor} \\
}
\startdata
$\Gamma=1.7$; N$_{H}=4\times10^{20}$ cm$^{-2}$ & 0.917 \\
$\Gamma=0.9$; N$_{H}=4\times10^{20}$ cm$^{-2}$ & 2.473 \\
$\Gamma=1.7$; N$_{H}=4\times10^{21}$ cm$^{-2}$ & 2.148 \\
$\Gamma=0.9$; N$_{H}=4\times10^{21}$ cm$^{-2}$ & 1.068 \\
\texttt{compmag} \citep{Ballhausen2017}& 1.44 \\
\enddata
\tablecomments{Conversion factors used to create lines for each spectral model in Figure \ref{xmm}. To create the lines, an array of 0.2-12.0 keV fluxes spanning the range of values shown in the figure were multiplied by the conversion factor to get the corresponding 4--25 keV flux for that model.}
\end{deluxetable*}



\end{document}